\documentclass[twocolumn]{aastex63}

\usepackage[utf8]{inputenc}
\usepackage[T1]{fontenc}
\usepackage{amsmath}
\usepackage{amsfonts}
\usepackage{amssymb}
\usepackage{graphicx}
\usepackage{grffile}
\usepackage{hyperref}
\usepackage[caption=false]{subfig}
\usepackage{array}
\usepackage{xcolor}
\usepackage{listings}
\usepackage{comment}
\usepackage{bm}
\usepackage{slashed}
\usepackage{mathtools}

\usepackage{siunitx}
\usepackage{multirow}
\usepackage{booktabs}
\usepackage[para]{threeparttable}
\usepackage{ulem}

\numberwithin{equation}{section}

\newcommand*{\inc}{\Delta}
\newcommand{\code}[1]{\texttt{#1}}
\newcommand*{\integrand}[1]{\text{d}#1}
\newcommand*{\ra}[1]{r_{\alpha}^{#1}}
\newcommand*{\rc}[1]{r_{m}^{#1}}
\newcommand*{\rx}[1]{r_{M}^{#1}}
\newcommand*{\racx}[3]{\ra{#1}\, \rc{#2}\, \rx{#3}}
\newcommand*{\rac}[2]{\ra{#1}\, \rc{#2}}
\newcommand*{\rax}[2]{\ra{#1}\, \rx{#2}}
\newcommand*{\rcx}[2]{\rc{#1}\, \rx{#2}}
\newcommand*{\rde}{r_{\inc E}}
\newcommand*{\erot}{E_{\rm rot}}
\newcommand*{\evib}{E_{\rm vib}}
\newcommand*{\eh}{E_{\rm H}}

\newcommand*{\KE}{\text{K.E.}}

\newcommand*{\rH}{\text{H}}
\newcommand*{\rD}{\text{D}}
\newcommand*{\rHt}{\rH_2}
\newcommand*{\dt}{\rD_2}

\DeclareSIUnit \atomicunit{a.u.}
\DeclareSIUnit \erg{erg}
\DeclareSIUnit \rydberg{Ry}
\DeclareSIUnit \epccm{\erg\per\centi\meter\cubed\per\second}
\DeclareSIUnit \clight{\text{\ensuremath{c}}}

\begin{document}

\title{Molecular Chemistry for Dark Matter}
\shorttitle{Dark Molecular Chemistry}

\author[0000-0002-0378-5195]{Michael Ryan}
\email{mzr55@psu.edu}
\affiliation{Institute for Gravitation and the Cosmos, The Pennsylvania State University, University Park, PA 16802, USA}
\affiliation{Department of Physics, The Pennsylvania State University, University Park, PA, 16802, USA}

\author[0000-0002-8677-1038]{James Gurian}
\email{jhg5248@psu.edu}
\affiliation{Institute for Gravitation and the Cosmos, The Pennsylvania State University, University Park, PA 16802, USA}
\affiliation{Department of Astronomy and Astrophysics, The Pennsylvania State University, University Park, PA, 16802, USA}

\author[0000-0002-6498-6812]{Sarah Shandera}
\email{ses47@psu.edu}
\affiliation{Institute for Gravitation and the Cosmos, The Pennsylvania State University, University Park, PA 16802, USA}
\affiliation{Department of Physics, The Pennsylvania State University, University Park, PA, 16802, USA}

\author[0000-0002-8434-979X]{Donghui Jeong}
\email{djeong@psu.edu}
\affiliation{Institute for Gravitation and the Cosmos, The Pennsylvania State University, University Park, PA 16802, USA}
\affiliation{Department of Astronomy and Astrophysics, The Pennsylvania State University, University Park, PA, 16802, USA}
\affiliation{School of Physics, Korea Institute for Advanced Study (KIAS), 85 Hoegiro, Dongdaemun-gu, Seoul, 02455, Republic of Korea}

\date{\today}
\begin{abstract}
Molecular cooling is essential for studying the formation of sub-structure of dissipative dark-matter halos that may host compact objects such as black holes. Here, we analyze the reaction rates relevant for the formation, dissociation, and transition of hydrogenic molecules  while allowing for different values of the physical parameters: the coupling constant, the proton mass, and the electron mass. For all cases, we re-scale the reaction rates for the standard molecular hydrogen, so our results are valid as long as the dark matter is weakly coupled and one of the fermions is much heavier than the other. These results will allow a robust numerical treatment of cosmic structure, in particular for mini-halos for which molecular cooling is important, in a dissipative dark matter scenario.
\end{abstract}


\section{Introduction} \label{sec:intro}
If the particle content of dark matter has enough complexity to support its own chemistry, the universe may contain a much richer array of structures than surveys of stars and galaxies have already revealed. Dark matter with chemistry can be dissipative, so that in sufficiently dense environments it will radiate kinetic energy and cool, allowing compact structures to form. While gravitational evidence on galactic scales indicates that any interactions among dark-matter particles must be somewhat weaker than those observed in the visible matter \citep{Markevitch:2003at}, some small-scale data may be better explained by dark matter with self interactions \citep{Bullock:2017xww}. Future observations of smaller-scale structure, including the possible compact-object detections by gravitational wave observatories, will further probe for dissipative dark-matter interactions \citep{Kouvaris2011,deLavallaz2010,Bramante2015,Bramante2014,Kouvaris2018,Shandera2018,Latif:2018kqv,Abbott:2018oah,Authors:2019qbw,Singh:2020wiq,Hippert2021} and may ultimately resolve the dark matter puzzle.  
    
As an illustrative and calculable example, we consider a dissipative dark matter model consisting of two oppositely charged fundamental fermions with masses $m$ and $M$, interacting via a $U(1)$ force mediated by a dark photon. 
The interaction strength is determined by a coupling constant, $\alpha$. We take $m\ll M$ and the dark photon to be massless. For simplicity, we assume that any non-gravitational interactions between dark matter and the Standard Model particles are too weak to be relevant. Here, Standard Model refers to the Standard Model of particle physics, but, hereafter, we shall use the term Standard Model for referring to atomic and molecular physics of just electrons and protons with Standard Model masses and electromagnetic interaction strength. Other dissipative scenarios, especially ``mirror" dark matter where the dark sector has a copy of the particle content of the Standard Model, have also been studied in the literature \citep{Berezhiani2001,Chang2019,Dessert2019,DAmico2017,Choquette2019}.

In the scenario we consider, pairs of oppositely charged particles ($M$ and $m$) can form bound states analogous to hydrogen atoms, we call them {\it dark atoms}, and a gas of these atoms can undergo all of the familiar electromagnetic processes, including energy-level transition, ionization, and recombination. In addition to the free-free (bremsstrahlung) emission from the ionized states, dark atoms can dissipate energy by radiating dark photons through recombination or collisional excitation followed by radiative decay \citep{Rosenberg2017, Buckley2018}. Because the neutral, bound state atoms play an important role, this dark matter model is often called ``atomic" dark matter \citep{goldberg_new_1986,Ackerman2009,Feng2009,Kaplan2010,Kaplan2011,Fan2013,CyrRacine2013,Cyr-Racine:2013fsa,Cline2014,Foot:2014uba,2016JCAP...07..013F,2015JCAP...09..057R,Agrawal:2016quu,Boddy2016,Ghalsasi:2017jna}. 

However, the energy dissipation allowed by atomic processes requires excited-state atoms and free electrons, which are rare at low gas temperature where essentially all atoms are neutral and in the ground state. So, atomic cooling is inefficient at low temperatures. In the Standard Model, for example, cooling below \SI{e4}{\kelvin} proceeds mostly through line cooling of molecular hydrogen \citep{Glover2012,Mo2010}. The lower-lying and more closely spaced molecular levels (see Sec.~\ref{sec:basic_tools}) allow the gas to cool to approximately \SI{100}{\kelvin}. The Jeans mass at this minimum temperature sets the mass scale of the collapsing gas cloud, which in the case of Pop.~III stars is correlated with the eventual mass of the proto-stars \citep{bromm1999, Bromm2002, Hirano2014}.

The molecular and atomic cooling processes for hydrogen are well-studied in the Standard Model. Of particular importance to us are the atomic and molecular processes for Population III star formation that take place within a pristine gas that is 92\% hydrogen.
In this paper we derive the chemical reaction rates and cooling functions relevant for the analogous processes in an atomic-dark-matter gas containing all species (free particles, atoms, ions, and molecules). Many of the atomic processes were derived in \citet{Rosenberg2017}, so our focus is on molecular physics.

The following three ingredients are necessary to understand the cooling processes of a gas of atoms and molecules \citep{Flower2007,Draine2011}: (1) the abundance of the gas comprising each species, (2) the complete set of quantum states available to each component of the gas and how those states are populated, and (3) the interaction potentials between each species. To obtain the information from first principles requires solving multi-particle Schr\"odinger equations. Even within the Standard Model the solutions to these equations, and the molecular states and scattering rates, can only be found approximately and numerically. The cooling functions used in simulations of structure formation are typically semi-analytic, and significant uncertainties in various rates remain. The state of the art is reviewed, for example, in \citet{GalliPalla2013, Glover2008, Glover2015a}.

Instead of obtaining the molecular wave functions, energies, and interactions from first principles, in this paper we model dark molecular processes based on the known results for Standard Model hydrogen. Fortunately, a little dimensional analysis goes a long way. For example, in the next section (Section \ref{sec:basic_tools}), we review the Born-Oppenheimer approximation for the quantum mechanical calculation of hydrogenic molecules and show how those results can be re-scaled by ratios of the dark matter parameter values to the Standard Model values:
\begin{align}
\label{eq:ratios}
    r_m=\frac{m}{\SI{511}{\kilo\electronvolt}}\,,
    r_M=\frac{M}{\SI{0.938}{\giga\electronvolt}}\,,
    r_{\alpha}=\frac{\alpha}{137^{-1}}\,.
\end{align}
We then argue that the re-scaling that we derive for the analytic but approximate Born-Oppenheimer treatment largely carries over to the more accurate but semi-analytic results used in the literature.\footnote{Appendix \ref{sec:mass_scaling} compares our re-scaling procedure to a simpler mass re-scaling technique used in literature on Standard Model reaction rates.} Sections \ref{sec:cooling} and \ref{sec:reactionrates} present the cooling functions and reaction rates obtained using these re-scaling techniques. Note that we assume $c=\hbar=1$ throughout except where noted.

The results we present in this work have a broad range of potential applications including a more accurate treatment of Standard Model deuterated chemistry. However, we developed the re-scaling scheme to compute two specific aspects of dissipative dark matter scenarios: (A) the cosmological evolution of the dark molecular hydrogen fraction and (B) formation of compact objects from dark molecular cooling. We apply the outcome of this paper to these cases in two companion papers, determining the early-time makeup of the cosmological gas in \citet{Gurian2021} and modeling the dark-matter halo cooling from both dark hydrogen atoms and molecules in \citet{Ryan2021}.

While many of the results in this paper are independent of cosmology, our eventual goal is to work out the observational consequences if some or all of the dark matter is atomic dark matter. For that purpose, we make the following four assumptions:
(A) dark charge neutrality, 
(B) sufficiently fast recombination for ``atomic" processes to be relevant, 
(C) thermalized particle distribution,
and
(D) collisional-ionization equilibrium. That is, we assume an equal number of $M$ and $m$ particles, as doing otherwise would result in long-range $U(1)$ interactions overwhelming the known bulk gravitational behavior to alter, for example, the Friedmann equation. We assume a recombination rate large enough that there is a cosmologically significant fraction of dark atoms. We undertake a detailed treatment of the cosmological recombination of atomic dark matter in a companion paper \citep{Gurian2021}. Briefly, Saha equilibrium implies that the dark atoms will recombine when the dark radiation temperature drops sufficiently far below the atomic binding energy, as long as the recombination rate exceeds the Hubble rate. This is the case for the entire range of parameters studied in our companion papers. Third, we assume the dark particles have approximately-Maxwellian velocity distributions at a common temperature. \citet{Agrawal2017} generically validate this assumption in the cosmological context. In the case of cooling atomic dark matter halos, \citet{Fan2013, Rosenberg2017} delineate a regime where equi-partition is maintained by elastic collisions with the dominant dark hydrogen species. We expect equi-partition must be maintained well beyond this regime by other mechanisms (i.e.~Landau damping \citep{Shu1992}). Fundamentally, as long as $\alpha \gg G M^2$,  the U(1) force will prevent the large scale charge-separations implied by differential cooling of heavy and light particles in a gravitational potential. Lastly, we calculate the cooling rates assuming collisional-ionization equilibrium and ignoring the effects of photons produced through cooling processes, like potential reabsorption or reionization. In the absence of strong radiation sources \citep{Wiersma/etal:2009}, this treatment is equivalent to assuming the optically thin gas where the photons escape the medium. This turns out to be a good approximation in the Standard Model \citep{Lykins/etal:2013}.

\section{Basic tools for dark molecular physics}
\label{sec:basic_tools}
The hydrogenic molecule is a system of two heavy particles of mass $M$ and charge $+1$, at positions ${\bf X}_A$, ${\bf X}_B$, and two light particles of mass $m$ and charge $-1$, at positions ${\bf x}_1$, ${\bf x}_2$, all interacting electromagnetically, with the fine-structure constant $\alpha$. The Schr\"odinger equation for stationary states of this system depends on the distance between each pair of particles (e.g., $x_{1A}$ between light particle 1 and heavy particle $A$) as
\begin{align}
\label{eq:H2Schrodinger}
H\Psi({\bf X}_A,{\bf X}_B;{\bf x}_1,{\bf x}_2)
= E\Psi({\bf X}_A,{\bf X}_B;{\bf x}_1,{\bf x}_2)\,,
\end{align}
with Hamiltonian
\begin{align}
H &=
-\frac{1}{2M}(\nabla^2_A+\nabla^2_B) -\frac{1}{2m}(\nabla^2_1+\nabla^2_2)\nonumber\\
&+\alpha\left(\frac{1}{X_{AB}}+\frac{1}{x_{12}}-\frac{1}{x_{1A}}-\frac{1}{x_{2A}}-\frac{1}{x_{1B}}-\frac{1}{x_{2B}}\right)\,.
\end{align}
Here, $\Psi({\bf X}_A,{\bf X}_B;{\bf x}_1,{\bf x}_2)$ is the wave function for the stationary state with energy $E$, and we use $\hbar=1$. We use the shorthanded notation for the distance $X_{AB}\equiv|{\bf X}_A-{\bf X}_B|$, and so on. 
As long as $M\gg m$, we can separate the Schr\"odinger equation into electronic part and nucleic parts by using the Born-Oppenheimer approximation as follows. First, an ansatz for the wave functions of the light particles (the electrons) is introduced assuming the heavy particles (the protons) remain stationary \citep{Bethe1997}. The electronic wave functions are only parametrically dependent on the nuclear separation $X_{AB}$ and can be used to find an effective potential between the two nuclei, which is in turn solved for the nucleic part of the wave functions.

The simplest ansatz for the electronic ground state is the symmetric combination of products of ground state wave functions $u_{is}$ for the individual hydrogen atoms:
\begin{equation}
        \psi = \frac{1}{\sqrt{2}} \left(u_{1s}(x_{1A})u_{1s}(x_{2B}) + u_{1s}(x_{2A})u_{1s}(x_{1B})\right)\,.
    \label{eq:HLWF}
\end{equation}
This is the Heitler-London \citep{Heitler27} wave function for the molecule, which depends on fundamental parameters only in the usual Bohr radius combination $a_0=(\alpha m)^{-1}$. This result is exact when the two hydrogen atoms are infinitely far apart, and provides the asymptotic form for more accurate electronic wave functions at large nuclear separation. Assuming this solution for the electronic wave functions, the effective potential for the nuclei (called the Heitler-London potential) can be written analytically. Its shape is very well approximated by the simpler Morse potential, a function of the nuclear separation:
\begin{align}
V_{\rm Morse}(X_{AB})=V_0(e^{-2(X_{AB}-X_0)/b}-2e^{-(X_{AB}-X_0)/b})\,.
\end{align}
where $X_0$ is the nuclear separation at the minimum of the potential. The typical scale for the parameters are $X_0, b \propto a_0$, $V_0\propto\alpha/a_0\propto m\alpha^2$.

The advantage of the Morse potential is that its energy levels can be found exactly, and analytic expressions for the corresponding nuclear wave functions can be obtained \citep{Dong2007}. The energies of the vibrational modes of the molecule are given approximately by the bound state energies in the Morse potential:
\begin{equation}
E_{{\rm vib},\nu}=-V_0\left(1-\frac{\nu+\frac{1}{2}}{K_0b}\right)^2\,,\;\;\; 0\leq \nu\leq K_0b-\frac{1}{2}\;,
\label{eq:Evib}
\end{equation}
where $K_0=\sqrt{2MV_0}$. Below the energy required to excite a vibrational mode, the separation of the two nuclei can be treated as fixed, $X_{AB}=X_0$, and then the Hamiltonian becomes that of a rigid rotor. By changing coordinates to those for the center of mass and relative motion, the usual rotational states are found with energy levels
\begin{equation}
E_{{\rm rot}, J} =\frac{J(J+1)}{MX_0^2}\,.
\label{eq:Erot}
\end{equation}

At this level of approximation, the molecular rotational energies and low-lying vibrational energies ($\frac{\nu}{K_0b}\ll1$) for generic masses $m\ll M$ and coupling ($\alpha\ll 1$) can be found from the Standard Model (SM) values via 
\begin{align}
\label{eq:rescaleEnergies}
E_{\bf mol.\,vib}&=\left[\frac{r_{\alpha}^2r_m^{3/2}}{r_M^{1/2}}\right]E_{\bf mol.\,vib, SM}\,,\nonumber\\
E_{\bf mol.\,rot}&=\left[\frac{r_{\alpha}^2r_m^{2}}{r_M}\right]E_{\bf mol.\,rot, SM}\,.
\end{align}

The literature on molecular cooling and scattering uses more accurate solutions for the molecular hydrogen energy eigenstates, found by introducing more sophisticated ansatze for the electronic wave functions. These are typically constructed from a basis set of trial wave functions, and then the task is to solve for the coefficients of the basis states that contribute to each stationary state. For example, the James-Coolidge basis \citep{James1933} for solving the Schr\"odinger equation is relatively simple and is valid for small inter-nuclear distances. In this basis, the ansatz for the electronic wave functions $u_{m, X_0}$ is written in terms of coordinates $y_{ij}\equiv x_{ij}/X_0$ in units of the fixed nuclear separation, which up to a constant is the same as units of the Bohr radius (e.g., $y_{1A}=x_{1A}/X_0\propto x_{1A}/a_0$):
\begin{align}
u_{m,X_0}
&=
\sum_{\{n\}}
c_{\{n\}}(1+P_{AB})(1+P_{12})
y_{12}^{n_1}
(y_{1A}-y_{1B})^{n_2}
\nonumber
\\
&\times
(y_{2A}-y_{2B})^{n_3}(y_{1A}+y_{1B})^{n_4}(y_{2A}+y_{2B})^{n_5}
\nonumber\\
&\times
e^{-\beta(y_{1A}+y_{1B})}
e^{-\beta(y_{2A}+y_{2B})}
\end{align}
where $P_{AB}$ ($P_{12}$) indicates the permutation of $A$ and $B$ (1 and 2), and the coefficients $c_{\{n\}}$ and $\beta$ are variational parameters. The individual indices $n_i$ are integers $\geq 0$, and the summation is over all sets of indices that satisfy $\sum_{i=1}^5 n_i\leq \Omega$ for $\Omega=3,4,5...$, up to the desired precision. This basis was used in \citet{Kolos1960} in an early determination of electronic wave functions, and more recently (e.g, \citet{Sims2006}) to make high-accuracy calculations for the hydrogen molecule. \citet{Pachucki2010} gives analytic expressions for many integrals needed in molecular calculations using these wave functions. Other classic literature, beginning with a paper by \citet{Kolos1965} generalizes the James-Coolidge basis to be more accurate at large inter-nuclear distances, where it should asymptote to the Heitler-London solution. But, none of these prescriptions introduce a new parametric dependence on the fundamental parameters into the molecular potential, and so the re-scaling shown in Eq.~(\ref{eq:rescaleEnergies}) should remain accurate.

Heuristically, we can obtain the re-scaling in Eq.~(\ref{eq:rescaleEnergies}) by using the {\it spring constant} for the vibrational mode
\begin{equation}
k 
= \left(\frac{{\rm d}^2V_{\rm Morse}}{{\rm d}X^2}\right)_{X=X_0}
= \frac{V_0}{b^2} \propto m^3 \alpha^4,
\end{equation}
which yields the vibrational energy level as
\begin{equation}
E_{\bf mol.\,vib} 
= \nu\hbar \omega 
= \nu\hbar \sqrt{\frac{k}{M}} \propto \sqrt{\frac{m}{M}} m\alpha^2\,,
\end{equation}
and the moment of inertia, $I=MX_0^2$, which yields the rotational energy level as 
\begin{equation}
E_{\bf mol.\,rot}
=
\frac{L^2}{2I}
\propto \frac{1}{Ma_0^2} 
= \frac{m^2\alpha^2}{M}\,.
\end{equation}
Note that these heuristic results rely on the facts that the typical length scale and energy scale for the hydrogen molecule are determined by the electronic states in the framework of Born-Oppenheimer approximation. That is, it is the Bohr radius and the hydrogen binding energy that set the relevant scales, and the heuristic re-scaling must hold beyond the specific electronic wave function being used. That argument further supports our extension of the re-scaling in Eq.~(\ref{eq:rescaleEnergies}).

The spin part of the nuclear wave functions is also important for understanding molecular properties. There is a singlet state, parahydrogen, with total nuclear spin $S_N=0$, and a triplet state, orthohydrogen, with total nuclear spin $S_N=1$. Since the total wave function must be antisymmetric under the exchange of the two heavy fermions (protons) the anti-symmetric singlet state (parahydrogen) can only have $J$-even rotational states, while the symmetric triplet state (orthohydrogen) must have $J$-odd. Because the majority of radiative transitions do not change the spin state, and a photon carries angular momentum $\Delta J=1$, the leading order radiative transitions of hydrogenic molecules are electric quadrupole transitions, which are slower compared to the dipole transition by a factor of $(a_0/\lambda_{u\ell})^2$ where $\lambda_{u\ell}$ is the wavelength corresponding to the energy gap.

\subsection{Molecular binding energy}
\label{sec:mol_bind_energy}
Molecular transitions and reaction rates generally depend on the ratio of the temperature to the energy scale associated with the reaction. For rotational and vibrational transitions, we presented the parametric dependence of the relevant energy scales in the previous section. On the other hand, formation and dissociation processes of molecular states involve a sum of partial cross sections over rovibrational states and so there is at least a formal dependence on multiple energy scales. Nonetheless, we argue that the most relevant energy scale for the formation and dissociation of hydrogenic molecules is the binding energy $\eh$ of hydrogen.

This is most obvious in the Heitler-London treatment of the problem. Writing down the symmetric combination of ground state wave functions for the individual hydrogen atoms, Eq.~(\ref{eq:HLWF}) leads to an expression for the expectation value of the total energy of the molecule of the form \citep{Heitler27, Dushman36}:
\begin{equation}
E_{\rHt} = \eh[1 + f(X_{AB}/a_0)],
\end{equation}
from which the binding energy is determined by minimizing the energy over $X_{AB}$ and comparing to the total energy of the separated hydrogen energies. There are two key observations. First, although molecular dissociation involves separating the nuclei, the binding energy is determined principally by the electronic configuration. Second, this binding energy is (at lowest order in perturbation theory) just a constant fraction of $\eh$, and the constant does not depend on mass ($m$, $M$) nor coupling constant $\alpha$. This method calculates a binding energy of $\SI{3.2}{\electronvolt}$ \citep{Heitler27} compared to the measured value of $\SI{4.74}{\electronvolt}$, and the same approach applied to $\rHt^+$ yields an energy of $\SI{1.77}{\electronvolt}$  compared to the true value of $\SI{2.64}{\electronvolt}$ \citep{Li2007}.  

Unfortunately, the accuracy of this method is clearly less than excellent. This is because the molecule is too tightly coupled for the Heitler-London wave function to be a good approximation to the molecular electronic wave function. Historically, this problem was first addressed by attaching some ansatz together with a variational parameter to the individual electron wave functions. One early such attempts is \citet{Rosen1931}. There, the individual electronic wave functions are given as for example
\begin{equation}
\psi_{1A} = u_{1s}(x_{1A}) + \sigma u'({\bf x}_{1A}),
\end{equation}
where $u_{1s}$ is the hydrogenic wavefunction for an effective (screened) nuclear charge $Z$, and $u'$ is the the wave function of the $2p$ state in a hydrogenic atom with charge $2Z$. Both $Z$ and $\sigma$ are determined variationally.  The full electronic wavefunction is taken as the Heitler-London combination of the individual electronic wavefunctions. This approach ultimately yields a ground state energy of $\SI{4.02}{\electronvolt}$, which differs by 15.1\% from the actual value, and again clearly scales as $\eh$.
     
To achieve percent-level accuracy requires a fully variational calculation, while also including the previously neglected electron-electron separation, in the James-Coolidge basis \citep{James1933} (a calculation which \citet{James1933} note ``can be evaluated and checked by an experienced computer in a little over two hours.''). This variational calculation may well contain some subdominant higher order energy parameter dependence which cannot be easily extracted. Still, we can be encouraged by two facts. First, the calculation is still concerned entirely with determining the electronic wave function as a function of fixed internuclear separation (and then minimizing the energy of that state with respect to the separation). Second, the variational ansatz carries dimensions only through the Bohr radius. All this makes clear that the molecular dissociation energy scales dominantly with the atomic binding energy, and that $T/\eh$ is the ``dimensionless temperature'' relevant to these reactions. Experimental data suggests that this conclusion is correct: the dissociation energy of $\rHt$ differs from that of $\dt$ by less than 2\% \citep{Herzberg61}, consistent with predominantly $\eh$ dependence, insensitive to the mass $M$. 

\subsection{Spontaneous emission rate}
\label{sec:emission}
Next, we consider the spontaneous emission rates, or Einstein $A$ coefficients. $\rHt$ is a homonuclear molecule with no permanent electric dipole moment, so transitions between states are quadrupolar, and the hard work is to compute the quadrupole transition $\langle \Psi_{\nu,J}|\hat{Q}|\Psi_{\nu^{\prime},J+2}\rangle$. Here, we bypass this calculation by re-scaling known results from the Standard Model to estimate the $A$ coefficient for the dark-atomic model.

The electric quadrupole radiation power is proportional to 
$P\propto\omega^6Q^2$ \citep{Jackson}, so equating that with $A^{\rm quad.}\hbar\omega$ we find the transition rate between states separated by energy $\Delta E$ scales as\footnote{In the literature, this rate is often given in atomic units. See Appendix \ref{app:AU} for a derivation of the re-scaling in those units.}
\begin{align}
    A^{\rm quad.}&\propto\frac{\alpha(\Delta E)^5}{\hbar^5c^4}\langle Q\rangle^2\,.
\end{align}
The literature contains at least two definitions for the quadrupole moment that differ by a factor of 2 \citep{Poll1978, Aannestad1973}, but $Q\propto {\rm (length)}^2\propto a_0^2$ regardless. In the equation above we have explicitly written the factors of $\hbar$ so that the units on the right hand side are obviously $\rm{s}^{-1}$. 

Then, the dark matter Einstein $A$ coefficient for transitions between rotational levels can be estimated by
\begin{align}
A_{\rm rot,\,DM}
\propto&\,
\alpha E_{\rm\bf mol.~rot}^5 (a_0^2)^2 \\
=&
r_{\alpha}\left[\frac{r_{\alpha}^2r_m^2}{r_M}\right]^5\left(\frac{1}{(r_{\alpha}r_m)^2}\right)^2\; A_{\rm rot,\;SM}\nonumber\\
=&\left[\frac{r_{\alpha}^7r_m^6}{r_M^5}\right]\; A_{\rm rot,\;SM}\,.
\label{eq:Aquad_rot}
\end{align}

Similarly, for transitions involving only the vibrational level, we estimate the rate by
\begin{align}
A_{\rm vib,\, DM}
\propto&\,
\alpha E_{\rm\bf mol.~vib}^5 (a_0^2)^2 \\
=&
r_{\alpha}\left[\frac{r_{\alpha}^2r_m^{3/2}}{r_M^{1/2}}\right]^5\left(\frac{1}{(r_{\alpha}r_m)^2}\right)^2\; A_{\rm vib,\;SM}\nonumber\\
=&\left[\frac{r_{\alpha}^7r_m^{7/2}}{r_M^{5/2}}\right]\; A_{\rm vib,\;SM}\,.
\label{eq:Aquad_vib}
\end{align}

\subsection{Collisional excitation scattering rates}
\label{sec:scattering}
In the simplest scattering between hydrogen molecules ($\rHt$-$\rHt$), or between a hydrogen molecule and a hydrogen atom ($\rHt$-$\rH$), there is a change in the state of the molecule but no exchange of particles, generally called collisional excitation. Here, we discuss how to re-scale the rates for molecular scattering. We then generalize the tools developed here to general scattering processes involved in molecular chemical reactions in Section \ref{sec:reactionrates}. 

With molecular states and energies in hand from Eq.~(\ref{eq:rescaleEnergies}), the additional ingredient needed for scattering rates is the interaction potentials between two molecules, or between a molecule and a hydrogen atom. \citet{Takayanagi1960} used a Morse-type potential in a classic analytic treatment of $\rHt$-$\rH$ scattering. The scattering rate coefficients, $\gamma\equiv\left<\sigma v\right>$, are determined by the velocity-averaged cross section $\sigma$ for changing the rotational states, where velocity $v$ is assumed to be drawn from a Maxwell-Boltzmann distribution at temperature $T$. \citet{Takayanagi1960} have denoted
\begin{align}
\gamma_{J,J+2}=&\langle \sigma v\rangle_{J\rightarrow J+2} \nonumber\\
=&\sqrt{\frac{8}{\pi\mu k_B^3  T^3}} \int_0^{\infty} \sigma_{J\rightarrow J+2} (E^{\prime})e^{-E^{\prime}/k_BT}E^{\prime}dE^{\prime}\nonumber\\
=&\sqrt{\frac{8 k_BT}{\pi \mu}}\pi R_c^2\langle F\rangle
\label{eq:approxHH2}
\end{align}
where $E^{\prime}$ is the center of mass collision energy, $\mu$ is the reduced mass of the $\rHt-\rH$ pair, $R_c\propto a_0$ is the distance of closest approach in a classical head-on collision. Assuming the Morse-type potential, $\langle F\rangle$ is the dimensionless result of an integral and is known analytically. Although this is a simple approach, it remains reasonably accurate even as calculations for the interaction potential have continued to improve. For example, \citet{Wrathmall2007} compared a Morse-type potential, although not identical to the one used in \citet{Takayanagi1960}, to a very accurate, recent $\rHt$-$\rH$ potential and found reasonable agreement. So, we will use the Eq.(\ref{eq:approxHH2}) to re-scale the scattering rates.

The dimensionality of $\gamma$ comes from the two factors of length in the cross section ($R_c\propto a_0$) and the thermal/kinematic factor from the velocity average, $\sqrt{k_BT/\mu}$. Since the energy difference between the initial and final levels remains parametrically important after the integration, it is useful to define the dimensionless combination $y^2=\inc E/(k_BT)$. While $\langle F\rangle$ is dimensionless, it does become a function of $y^2$, due, at minimum, to the  presence of $T$ in the exponential term. Then, if a scattering rate is known to scale with temperature as $f(T)=f(\inc E/y^2)=f_{\inc E}(y)$, and using the dependence of $R_C\propto a_0$ on fundamental parameters and $\mu\propto M$, the complete parameteric dependence is 
\begin{equation}
\label{eq:rescale_general}
\gamma\propto\sqrt{\frac{\inc E}{M}}\frac{1}{\alpha^2m^2}f_{\inc E}(y).
\end{equation}

Collisions in general cause transition between both vibrational and rotational states. At each gas temperature, however, we can approximate the collisional transitions by using the dominant transition: Cooling via de-excitation of rotational levels in the vibrational ground state ($\nu=0$) dominates at lower temperatures, and cooling via transitions between the lowest few vibrational levels (averaging over the more closely spaced rotational levels) dominates at higher temperatures.

In this approximation, we first consider transitions between rotational levels within a fixed vibrational level. Let us define the dimensionless parameter
\begin{equation}
y^2_{\rm rot}=\frac{\inc \erot}{k_B\,T}=\frac{m^2\alpha^2}{M k_B\,T}\,.
\end{equation}
Then, the rotational transition rate may be written as
\begin{align}
\gamma\propto&
\frac{1}{\alpha^2m^2}
\sqrt\frac{\inc \erot}{M}f_{\rm rot}(y_{\rm rot}) \nonumber\\
\propto&
\frac{1}{\alpha^2m^2}
\sqrt\frac{\alpha^2m^2}{M^2}f_{\rm rot}(y_{\rm rot}) \nonumber\\
=&
\frac{f_{\rm rot}(y_{\rm rot})}{\alpha m M}\,.
\label{eq:gamma_rot_parameter_scale}
\end{align}
The re-scaling for the dimensionful prefactor in the dark-matter model rate coefficient is then obvious from Eq.~(\ref{eq:gamma_rot_parameter_scale}). We can think of $y_{\rm rot}^2$ as the temperature in $\inc \erot$ units, the relevant energy scale of the problem. For a given dark-matter temperature $T_{\rm DM}$, we have to evaluate the function $f_{\rm rot}$ at the corresponding Standard Model temperature $T_{\rm SM}$ yielding the same $y^2_{\rm rot}$:
\begin{equation}
    T_{\rm SM} = T_{\rm DM}\;\left(\frac{\inc E_{\rm rot, SM}}{\inc E_{\rm rot, DM}}\right).
\end{equation}
Using the rotational-energy re-scaling in Eq.~(\ref{eq:rescaleEnergies}), then, we can define a re-scaled temperature, 
\begin{align}
\label{eq:Trscale}
\tilde{T}_r\equiv &\left(\frac{r_{M}}{r_{m}^2r_{\alpha}^2}\right)\;T,
\end{align}
and obtain the final dark matter rate coefficient as 
\begin{align} 
	\gamma_{J,{\rm DM}}(T)\approx&\left[\racx{-1}{-1}{-1}\right]\; \gamma_{J, SM}(\tilde{T}_r)\,.
	\label{eq:basic_gamma_rot_scale}
\end{align}

The considerations for molecule-molecule scattering are more complicated than the molecular-atomic case, as the para-para, ortho-ortho, and ortho-para cases must all be considered \citep{McMahan1974,Silvera1980}. In addition, the relative orientations of the molecules is important.
Still, it seems reasonable that the basic re-scaling for molecule-molecule scattering should be similar to the molecule-atom case: the cross section should increase as the size of the molecule ($R_c\propto a_0$) increases, and the reduced mass in the denominator be nearly $M$. We therefore use Eq.~(\ref{eq:basic_gamma_rot_scale}) to re-scale rates involving a change in rotational level for both $\rHt$-$\rHt$ and $\rHt$-$\rH$ processes. 

For collisions involving a change in the vibrational level, the calculation proceeds along the same lines: the wave functions for the initial and final states are different from above, but the interaction potential is the same. So, defining 
\begin{equation}
y^2_{\rm vib}=\frac{m^{3/2}\alpha^2}{M^{1/2}k_BT}\,,
\end{equation}
assuming the collisional excitation rate similarly contains a dimensionless function that now depends on $y_{\rm vib}$, $f_{\rm vib}(y_{\rm vib})$, and including the remaining terms from Eq.~(\ref{eq:rescale_general}), we find that 
\begin{align}
\gamma_{\rm vib}\propto&\frac{1}{\alpha^2m^2}\sqrt{\frac{\Delta E_{\rm vib}}{M}}
f_{\rm vib}\left(y_{\rm vib}\right) \nonumber\\
\propto& 
\frac{1}{\alpha^2m^2}
\sqrt\frac{m^{3/2}\alpha^2}{M^{3/2}}
f_{\rm vib}\left(y_{\rm vib}\right) \nonumber\\
=&
\frac{f_{\rm vib}(y_{\rm vib})}{\alpha\, m^{5/4}\,M^{3/4}}\,,
\label{eq:rescalevgeneral}
\end{align}
for which the re-scaling for the atomic dark-matter becomes
\begin{align}
	\gamma_{\rm vib, DM}(T)\approx& \left[\racx{-1}{-5/4}{-3/4}\right]\; \gamma_{\rm vib, SM}(\tilde{T}_v)\,,
\label{eq:rescale_gamma_vib}
\end{align}
with
\begin{equation}
\label{eq:Tvscale}
\tilde{T}_v\equiv \left(\frac{r_{M}^{1/2}}{r_{m}^{3/2}r_{\alpha}^2}\right)\,T.
\end{equation}
Again, we apply this formula for vibrational transition rates from both $\rHt$-$\rHt$ and $\rHt$-$\rH$ collisions.

\subsection{Summary}
\begin{table*}[h!t]
	\centering
	\footnotesize
	\begin{tabular}{l c r}
			\toprule
			Quantity & Dependence & Re-scaling \\
			\midrule
			$a_0$ (Bohr radius) & $\frac{1}{\alpha\,m }$ & $r_{\alpha}^{-1} \, r_m^{-1}$\\
			$\eh$ (atomic energy level spacing) & $m \, \alpha^2$ & $r_{\alpha}^{2} \, r_m$\\
			$\erot$ (Molecular rotational energy) & $\frac{\alpha^2 \, m^2}{M}$ & $\racx{2}{2}{-1}$\\
			$\evib$ (Molecular vibrational energy) & $\frac{\alpha^2\, m^{3/2}}{M^{1/2}}$  & $\racx{2}{3/2}{-1/2}$\\
			${\bf d}$ (dipole moment) & $0$  & $0$\\
			$A_{\rm rot}$ (quadrupolar rotational Einstein coefficient) & $ \frac{\alpha^7\,m^6}{M^5}$ & $\racx{7}{6}{-5}$\\
			$A_{\rm vib}$ (quadrupolar vibrational Einstein coefficient) & $\frac{\alpha^{7}\, m^{7/2}}{M^{5/2}}$ & $\racx{7}{7/2}{-5/2}$ \\
			$p_{ij}$ (polarizability) & $a_0^3 = \frac{1}{m^3\,\alpha^3}$ & $r_{\alpha}^{-3} \, r_m^{-3}$\\
			\bottomrule
		\end{tabular}
		\caption{Table of primary quantities important for reaction rates, their dependence on the parameters $m$, $M$, and $\alpha$, and the powers of ratios of parameters (see Eq.(\ref{eq:ratios})) needed to re-scale the quantities.
		\label{tab:base_quantity_scaling}}
	\end{table*}
We can extract the parametric dependence of many relevant processes in the atomic-dark-matter model by re-scaling. As we have shown, reaction and transition rates can at lowest order be treated as dependent only on some length scale (proportional to $a_0$) and some energy scale (atomic,  rotational, or vibrational). Since $\rHt$ has no dipole moment (${\bf d}=0$), $\rHt$ state transitions are quadrupolar, with the vibrational and rotational Einstein coefficients derived above. We neglect higher moments, which are less important for scattering rates and cooling processes. Lastly, several reaction rates will require the ground state polarizability, a tensor mediating the external electric field and the induced electric dipole moment, $p_{ij}\propto a_0^3 \propto 1/(m^3 \, \alpha^3)$. This is straightforward to show for the $\rH$ atom (see e.g. \citet{Sakurai2017} for a discussion using perturbation theory), but is more complicated for the $\rHt$ molecule \citep{Kolos1967}.  We have listed the parametric dependence and re-scaling equation for all of the aforementioned quantities in Table \ref{tab:base_quantity_scaling}.

\section{Application: Rovibrational cooling for dark molecular hydrogen}
\label{sec:cooling}
Molecular hydrogen cooling becomes important at low temperatures and high $\rHt$ concentrations in the Standard Model \citep{Glover2012}. In these conditions, molecular line cooling is the dominant process, with certain $\rHt$ production and destruction channels providing auxiliary cooling. In this section, we calculate the molecular line cooling rate for the atomic-dark-matter model by applying the re-scaling that we have derived in Section~\ref{sec:emission}-\ref{sec:scattering}. To start, we briefly derive the molecular line cooling equations in Sections \ref{sec:mol_line_cool} and \ref{sec:mlc_multi}, connecting our notation to that in the classic paper of \citet{Hollenbach1979}, whose analytic model forms the basis of comparison with the more modern computational and empirical results \citep{Glover2008,Glover2015,Galli1998} that we would prefer to use. In Sections \ref{sec:gen_rotation} and \ref{sec:gen_vibration} we re-scale the equations based on the dominant physical process, and in Section \ref{sec:rovib_cooling_scaling} we combine the equations to obtain the full, re-scaled, rovibrational cooling rates. We present full results based on \citet{Hollenbach1979} as well as \citet{Glover2008}. Since it is useful to gain an understanding of the rates using the analytic results of \citet{Hollenbach1979}, who considered only on $\rHt-\rH$ and $\rHt-\rHt$ collisions, Sections \ref{sec:gen_rotation} -\ref{sec:rovib_cooling_scaling} concentrate on those as collision partners. However, at low particle densities molecular line cooling can be driven by collisions with various other species \citep{Glover2008}, which we discuss in Section \ref{sec:other_species}.

\subsection{Molecular line cooling}
\label{sec:mol_line_cool}
Molecular line cooling occurs when the {\it coolant} molecules in excited states decay to lower-level states by spontaneously emitting photons. The cooling rate per volume (in units of \si{\erg\per\second\per\centi\meter\cubed}) can be written as the sum over all radiative transitions: 
\begin{equation}
\label{eq:Ccool}
{\cal C} = \sum_{u} n_u \sum_{\ell<{u}} A_{u\ell}\inc E_{u\ell}\,,
\end{equation}
where $n_u$ is the number density of molecules in the upper-energy state $u$, the $\ell$ are the lower-energy states available for the transition,  $A_{u\ell}$ is the Einstein $A$ coefficient, and $\inc E_{u\ell} = E_u-E_\ell$ is the energy carried away by the photon. 

Evaluating the re-scaled cooling rate from Eq.~(\ref{eq:Ccool}) requires knowing three pieces of information: the occupation numbers of the various rotational/vibrational levels, the re-scaled Einstein $A$ coefficient, and the re-scaled energy of each eigenstate. We have already derived the energies in Section \ref{sec:basic_tools}, so the energy differences are given approximately as
\begin{align}
&\inc E(\nu',J\pm2,\nu,J) \nonumber\\
=\,& (\evib(\nu')-\evib(\nu)) 
 +(\erot(J\pm2)-\erot(J)),
\end{align}
where the vibrational and rotational energies are approximated by Eq.~(\ref{eq:rescaleEnergies}). We have also derived the re-scaling of the Einstein $A$ coefficient for the electric quadrupole transition in Eqs.~(\ref{eq:Aquad_rot})-(\ref{eq:Aquad_vib}).
The remaining quantity to compute is the number density of molecules in the excited states, which depends on the temperature and the collision rate with the colliding particle $X$. 

To simplify the discussion, let us first consider the simple two-level system. If a gas is in (local) thermal equilibrium (LTE) at temperature $T_{\rm gas}$, then the ratio of the population of the two states is given as the Boltzmann factor:
\begin{equation}
\frac{n_u^{\rm LTE}}{n_\ell^{\rm LTE}}=\frac{g_u}{g_\ell}\exp (-\Delta E_{u\ell}/k_BT_{\rm gas})\,,
\label{eq:two_level_LTE}
\end{equation}
where $g_{u,\ell}$ are the number of internal degrees of freedom for each state. Considering those same levels as a two-state system subject to collisional excitation, collisional de-excitation, and radiative decay (without ambient radiation, or ignoring the stimulated emission or absorption), then the population of the excited state changes according to
\begin{equation}
\frac{dn_u}{dt}=n_Xn_\ell\gamma_{\ell u}-n_Xn_u\gamma_{u\ell}-n_uA_{u\ell}
	\label{eq:two_state_density_differential}
\end{equation}
where $n_X$ is the number density of the colliding species and $\gamma_{\ell u}$ ($\gamma_{u\ell}$) is the rate coefficient for the collisional excitation (de-exitation).
The steady state solution ($dn_u/dt=0)$ is
\begin{equation}
\label{eq:2levSS}
\frac{n_u}{n_\ell}
=\frac{n_X \gamma_{\ell u}}{n_X\gamma_{u\ell}+A_{u\ell}}\,.
\end{equation}
There are two limiting cases. 
First, when the radiative decay dominates over the collisional de-excitation, ($A_{u\ell}\gg n_X\gamma_{u\ell}$), the population density in the excited states is given as 
$n_u = n_\ell n_X\gamma_{\ell u}/A_{u\ell}$, or the collisionally excited molecules stay in the excited states for the mean lifetime of $A_{u\ell}^{-1}$.
On the other hand, when the collisional de-excitation dominates over the radiative decay, ($n_X\gamma_{u\ell}\gg A_{u\ell}$), the population at the excited state returns to the thermal solution in Eq.~(\ref{eq:two_level_LTE}) because 
$n_u
=n_\ell\gamma_{\ell u}/\gamma_{u\ell}
=n_\ell g_u/g_\ell \exp\left(-\Delta E_{u\ell}/k_BT_{\rm gas}\right)$ from the principal of detailed balance. These limits are commonly denoted as the $n\rightarrow0$ (since $n_X \ll A_{u\ell}/\gamma_{u\ell}$) and the $\rm LTE$ cases, respectively.

For a general system with more than two levels, collisional excitation, de-excitation and radiative decay can happen between any pairs of levels, so the equilibrium condition for the $i$-state becomes
\begin{equation}
\label{eq:nlevel_master}
\sum_{j\neq i} n_jn_X \gamma_{ji}
+
\sum_{j>i} n_j A_{ji}
=
\sum_{j\neq i} n_in_X \gamma_{ij}
+
\sum_{j<i} n_i A_{ij}\,.
\end{equation}
The solutions for the three-level case are given in Section 17.5 of \citet{Draine2011}. For systems with more than three levels, the equations are typically solved numerically. 

As the number density of the colliding particle determines the collisional de-excitation rate, it is useful to define a critical density for level $u$ of a collision partner $X$, $n_{\rm crit, u}(X)$, as a benchmark for when collisional de-excitation is important for determining the excited-level population. Although different definitions exist, the one used in \citet{Draine2011}, for example, (in the limit of no ambient radiation) is
\begin{equation}
\label{eq:ncrit}
n_{\rm crit, u}(X)\equiv\frac{\sum_{\ell<u}A_{u\ell}}{\sum_{\ell<u}\gamma_{u\ell}(X)}\,.
\end{equation}
That is, this is the density of a collision particle for which collisional de-excitation equals radiative de-excitation.

The relationship between the cooling rate in Eq.~(\ref{eq:Ccool}) and the cooling functions $L$ used in \citet{Hollenbach1979}, specific to a particular colliding particle $X$, is:
\begin{align}
\label{eq:LHM}
L_{u\ell}^{X}
=&\frac{1}{n(X)}\frac{1}{n(\rHt)}
\sum_{u} n_u \sum_{\ell<u} A_{u\ell}\inc E_{u\ell}\nonumber\\
=&\frac{{\cal C}}{n(X)n(\rHt)}\,,
\end{align}
where $L$ has units of \si{\erg\centi\meter\cubed\per\second}. Note that this definition is motivated by the fact that the upper-level population is proportional to the density of the colliding particle $n_X$ in the low-density limit, $n_X<n_{\rm crit, u}(X)$. For example, for a simple two-level system, 
${\cal C}= n_\ell n_X\gamma_{\ell u}\Delta E_{u\ell}$ at the low-density limit, $n_X\ll n_{\rm crit}$,
and 
${\cal C}= n_uA_{u\ell}\Delta E_{u\ell}$ at the high-density limit, $n_X \gg n_{\rm crit}$.

\subsection{Molecular line cooling with multiple colliding particles}
\label{sec:mlc_multi}
In the Standard Model, atomic hydrogen is the main collisional partner of molecular hydrogen, simply due to the high abundances, but several other species are also relevant for cooling, including other hydrogen molecules and helium.. When there is more than one colliding species then, analogous to Eq.~(\ref{eq:2levSS}), the steady-state population of the upper state (in the two-level approximation where we assume every pair of states separately satisfy the stationary condition) may be approximated as
\begin{equation}\label{eq:multi-population}
\frac{n_u}{n_\ell}
=
\frac{\sum_X n_X\gamma_{\ell u}(X)}
{\sum_X n_X \gamma_{u\ell}(X) + A_{u\ell}}\,,
\end{equation}
with which the cooling rate can be written as
\begin{equation}
{\cal C}
=
\sum_u 
\sum_{\ell<u}
\frac{\sum_X n_X\gamma_{\ell u}(X)}
{\sum_X n_X \gamma_{u\ell}(X) + A_{u\ell}}
n_\ell A_{u\ell} \Delta E_{u\ell}\,.
\label{eq:C_full_twolevel}
\end{equation}
Let us begin by considering the two limiting cases for the density of colliding partners: the low-density and high-density limits.

When the density of colliding particles is small ($n\to0$ limit) so that the de-excitation happens only radiatively, the cooling rate becomes,
\begin{equation}
{\cal C}_{n\to0}
\equiv
\sum_X
{\cal C}_{n\to0}^{X}
=
\sum_{X} 
\sum_{u}\sum_{\ell<u}
n_X 
n_\ell 
\gamma_{\ell u}(X) 
\Delta E_{u\ell}\,,
\label{eq:def_CXn0}
\end{equation}
or every collision leads to an emission of photon with energy $\Delta E_{u\ell}$.

In the opposite ($n\to\infty$) limit, we recover the LTE cooling rate as follows. Because the $\rHt$ molecules must be in the LTE state when the collision is frequent enough, and this must be true even with a single colliding particle species, the collisonal excitation rate and the collisional de-excitation rate must satisfy individually the detailed balance relation:
\begin{equation}
\frac{\gamma_{\ell u}(X)}{\gamma_{u\ell}(X)}
=
\frac{n_u^{\rm LTE}}{n_\ell^{\rm LTE}}
=
\frac{g_u}{g_{\ell}}
e^{-E_{u\ell}/k_BT}\,.
\label{eq:gamma_ratio}
\end{equation}
That means when collisional de-excitation dominates over the radiative decay,
\begin{equation}
A_{u\ell} \ll \sum_X n_X\gamma_{u\ell}(X)\,,
\end{equation}
we recover the LTE cooling rate
\begin{align}
{\cal C}_{\rm LTE}
=&\,
\sum_{u}
n_u
\sum_{\ell<u}
A_{u\ell}\Delta E_{u\ell}
\nonumber\\
=&\,
\frac{n(\rHt)}{\cal Z}
\sum_{u}
g_u e^{-E_{u}/k_BT}
\sum_{\ell<u}
A_{u\ell}\Delta E_{u\ell}\,,
\label{eq:C_LTE}
\end{align}
where
$n_u = (n(\rHt)/{\cal Z})g_u e^{-E_{u}/k_BT}$
with the partition function
${\cal Z} = \sum_i g_i e^{-E_i/k_BT}$.

In the intermediate density regime, rather than fully solving for the level population as a function of density of all colliding particles [Eq.~(\ref{eq:nlevel_master}) or its two-level approximation in Eq.~(\ref{eq:multi-population})], 
it is a common practice, for example in \citet{Hollenbach1979,Galli1998,Glover2008}, 
to interpolate between the low-density and high-density (LTE) cases by
\begin{align}
{\cal C} \simeq \frac{{\cal C}_{\rm LTE}}{1+{\cal C}_{\rm LTE}/\sum_X {\cal C}_{n\to 0}^X}\,,
\label{eq:density_interpolation}
\end{align}
so that ${\cal C}$ approaches to the correct limit at both low-density and high-density limits.

To understand this approximation better, let us explore the case when the approximation becomes exact. Starting from the two-level approximation in Eq.~(\ref{eq:C_full_twolevel}), and using 
Eq.~(\ref{eq:gamma_ratio}), we find the cooling rate as
\begin{align}
{\cal C}
=&
\sum_u 
\sum_{\ell<u}
\frac{\sum_X n_X\gamma_{\ell u}(X)}
{\sum_X n_X \gamma_{u\ell}(X) + A_{u\ell}}
n_\ell A_{u\ell} \Delta E_{u\ell}
\nonumber
\\
=&
\sum_u 
\sum_{\ell<u}
\frac{[n_u^{\rm LTE}/n_\ell^{\rm LTE}] n_\ell/n_u}
{1+ A_{u\ell}/\sum_X[n_X \gamma_{u\ell }(X)]}
n_u A_{u\ell} \Delta E_{u\ell}\,.
\end{align}
That is, in two-level approximation, the interpolation in Eq.~(\ref{eq:density_interpolation}) is exact if the level population follows the LTE value, for which 
\begin{align}
\frac{A_{u\ell}}{\sum_X[n_X \gamma_{u\ell}(X)]}
=&\,
\frac{A_{u\ell}n_u\Delta E_{u\ell}}
{\sum_X[n_X \gamma_{u\ell}(X) n_u\Delta E_{u\ell}]}
\nonumber\\
=&\,
\frac{{\cal C}_{\rm LTE}^{\rm 2-level,\,LTE}}{{\cal C}_{n\to 0}^{\rm 2-level,\,LTE}}\,.
\end{align}
In reality, this approximation may work better for $T<\Delta E$ so that only a few lowest-level states are populated. This is particularly true for the rotational state where $\Delta E\propto J$, but may be little worse for the vibrational modes where the energy gap stays constant.
%
%
%
%

\subsection{Rotational cooling rates} \label{sec:gen_rotation}
We must also account for the type of energy transition (rotational versus vibrational), separate from colliding particle density, since the results of Section \ref{sec:basic_tools} have demonstrated they have different re-scalings.  We first consider rotational transitions between levels $J$ and $J-2$, in the vibrational ground state $\nu=0$.
Let $\gamma_{JJ'}^{\rH}(T)$, $\gamma_{JJ'}^{\rHt}(T)$  be the rate coefficient for the ($J\to J'$) transition to occur due to a collision with, respectively, a hydrogen atom and molecule. 

At the low-density ($n\to0$) limit, with much suppressed collisional excitation, only the ground states (of both the para, $J=0$, and ortho, $J=1$, type molecules) are populated, and the dominant excitations are to the $J=2$ and $J=3$ states. Beginning with Eq.~(\ref{eq:Ccool}) and Eq.~(\ref{eq:LHM}), we have
\begin{align}
\label{eq:rotn0}
	&{\cal C}_{\rm{rot},n\rightarrow0}^{X}
	\nonumber\\
	=\,&n_{J=2}A_{20}\inc E_{20}+n_{J=3}A_{31}\inc E_{31}\nonumber\\
	=\,&
	n_{J=0}n(X)\gamma^X_{02}\inc E_{20} +n_{J=1}n(X)\gamma^X_{13}\inc E_{31}
	\nonumber\\
	=\,&
	\left(
	\frac{1}{4}\gamma^X_{02}\inc E_{20}
	+ \frac{3}{4}\gamma^X_{13}\inc E_{31}
	\right) n(\rHt)n(X)
	\nonumber\\
	=\,&
	\left(
	\frac{5}{4}\gamma^X_{20}e^{-\Delta E_{20}/k_BT}\inc E_{20}
	+ \frac{7}{4}\gamma^X_{31}e^{-\Delta E_{31}/k_BT}\inc E_{31}
	\right) 
	\nonumber\\
	&\times n(\rHt)n(X)\,,
\end{align}
where we use the low-density limit of Eq.~(\ref{eq:2levSS}), 
$n_u = (n_\ell n(X)\gamma_{\ell u})/A_{u\ell}$, to calculate the upper-level population, and set $n_{J=0}/n(\rHt)=1/4$, $n_{J=1}/n(\rHt)=3/4$ from the ortho-para ratio. The final equality follows from the detailed-balance relation in Eq.~(\ref{eq:gamma_ratio}). The total cooling rate is given as the sum ${\cal C} = \sum_{X=\rH,\rHt} {\cal C}^X$. Here we have assumed the commonly used ortho-para ratio of 3/1 and use that for the comparison of cooling rates in Fig.~(\ref{fig:ldl_scaled_cooling}). In general, however, the balance between the two spin states may be different. In Appendix \ref{app:orthopara} we derive the re-scaling for several processes that affect the ortho-para ratio and show how they depend on the dark parameters.

\citet{Hollenbach1979} provide an analytic form for a fit to the low-density limit collisional de-excitation rates 
\begin{eqnarray}
\gamma_{J,J-2}^{\rH}(T)
&=& \num{e-12} \left(\frac{10T_3^{1/2}}{1+60T_3^{-4}}+T_3\right)
\nonumber
\\&&
\hspace{-.8cm}
\times
\left(0.33+0.9\exp\left[-\left(\frac{J-3.5}{0.9}\right)^2\right]\right)\si{\centi\meter\cubed\per\second}
\nonumber
\\
\gamma_{J,J-2}^{\rHt}(T)
&=& \num{3.3e-12} \left(1+2T_3\right)
\nonumber
\\
&&
\hspace{-.8cm}
\times\left(0.276J^2\exp\left[-\left(\frac{J}{3.18}\right)^{1.7}\right]\right)\si{\centi\meter\cubed\per\second}
\label{eq:gammaJJm2}
\end{eqnarray}
where $T_3=T/1000~{\rm K}$. Because the $J$-dependent part in the second parenthesis is near unity for $J=3$, we can see that $\gamma^{\rHt}\gg \gamma^\rH$ for $T_3<1$. \citet{Hollenbach1979}  reported at the time that Eq.~(\ref{eq:gammaJJm2}) was $50~\%$ accurate for $0.1<T_3<5$, above which we start to see the vibrational line cooling, though subsequent results (e.g. \citep{Glover2008,Lee2008,Lique2015}) have shown the rate actually differs significantly at low temperatures. However, these rates, and the vibrational rates below, are still useful as an analytic comparison to demonstrate the basic re-scaling behavior.

Using our result in Eq.~(\ref{eq:basic_gamma_rot_scale}), the dependence on microphysical parameters can be captured by the re-scaling
\begin{align} 
\gamma_{J,J-2,{\rm DM}}^{\rH(\rHt)}(T)= \left[\racx{-1}{-1}{-1}\right]\;\gamma^{\rH(\rHt)}_{J,J-2, \rm{SM}}(\tilde{T}_r),
\end{align}
for which we evaluate the Standard Model rates at re-scaled temperature $\tilde{T}_r = r_M/(r_\alpha^2 r_m^2T)$.
Finally, substituting into Eq.~(\ref{eq:rotn0}) along with the energy re-scaling in Eq.~(\ref{eq:rescaleEnergies}), gives the re-scaling for the cooling rate
\begin{align}
\frac{{\cal C}_{{\rm rot},n\to0,\rm DM}^X(T)}{n(\rHt)n(X)}
=
\left[\racx{}{}{-2}\right] \; \frac{{\cal C}_{{\rm rot},n\to0,\rm DM}^X(\tilde{T}_r)}{n(\rHt)n(X)}
.
\label{eq:ldl_rot_scale}
\end{align}
	
At the high-density limit, collisional processes dominate so that all rotational levels are in thermal equilibrium, and the cooling rate is given by Eq.~(\ref{eq:C_LTE}):
\begin{align}
&{\cal C}_{\rm rot, LTE}(T,n(\rHt))
\nonumber\\
=&\,
n(\rHt)\sum_{J\ge2}
g_J
\frac{\exp[-E_J/k_BT]}{{\cal Z}(T)}A_{J,J-2}\inc E_{J,J-2}\,.
\end{align}
The re-scaling results from Section \ref{sec:basic_tools} for the Einstein $A$ coefficient [Eq.~(\ref{eq:Aquad_rot})] and the energy levels [Eq.~(\ref{eq:rescaleEnergies})] can be used to determine the re-scaled cooling rate in the LTE regime to account for changes in the microphysical parameters as following:
\begin{equation}
{\cal C}_{\rm{rot},\, LTE, DM}(T) 
=
\left[\racx{9}{8}{-6}\right] \;
{\cal C}_{\rm rot,\, LTE, SM}(\tilde{T}_r).
		\label{eq:lte_rot_scale}
\end{equation}

We then calculate the cooling rate in the intermediate density regime by interpolation following Eq.~(\ref{eq:density_interpolation}).

\subsection{Vibrational cooling rates} \label{sec:gen_vibration}
For cooling by vibrational transitions, we follow \citet{Hollenbach1979} who approximate the states as a three-level system ($\nu=2,1,0$), assuming all levels have the same number of internal degrees of freedom ($g_0=g_1=g_2=1$). We also follow them in approximating the vibrational motion as the simple harmonic oscillator, so that the energy levels are proportional to $\nu$ (dropping the terms higher order in $\nu$): $\inc E_{10}=\inc E_{21}=0.5\inc E_{20}$. Because the rotational energy gap is much smaller than the vibrational gap, we ignore any differences in energy if the $J$ value changes in the transition.

The low-density limit for this system comes from a similar calculation as was used above for the rotational case, except that the factor from degrees of freedom is one (compared to $(2J+1)/(2J-3)$) and there is no distinction between para and ortho states:
\begin{align}
\label{eq:vibn0}
&\frac{{\cal C}_{\rm{vib},n\rightarrow0}^{X}}{n(\rHt)n(X)\Delta E_{10}
}
\\
\nonumber
=&\,
\gamma_{10}^{X}e^{-\inc E_{10}/k_BT}
+2\gamma^{X}_{20}e^{-\inc E_{20}/k_BT}
+\gamma^{X}_{21}e^{-\inc E_{21}/k_BT}
\,,
\end{align}
which is nearly the expression given by \citet{Hollenbach1979} except that we have included the $\gamma_{21}$ term dropped there. The expression including the $\gamma_{21}$ term provides a better fit to the later results of \citet{Glover2008}. 
As for the collisional de-excitation rates, we use \citet{Hollenbach1979}\footnote{Although the $\gamma_{21}^{\rHt},\,\gamma_{10}^{\rHt}$ rate coefficients have been updated in \citet{Hollenbach1989}, we use the original here and in Figures \ref{fig:ldl_scaled_cooling} and \ref{fig:total_cooling_curve} for simplicity.}(rates are in \si{\centi\meter\cubed\per\second}):
\begin{eqnarray}
\gamma_{10}^{\rH} &=& \num{1.0e-12} T^{1/2} \exp\left[-\frac{1000}{T}\right] \\
\gamma_{20}^{\rH} &=& \num{1.6e-12} T^{1/2} \exp\left[-\left(\frac{400}{T}\right)^2\right] \\
\gamma_{21}^{\rH} &=& \num{4.5e-12} T^{1/2} \exp\left[-\left(\frac{500}{T}\right)^2\right] \\
\gamma_{10}^{\rHt} &=& \num{1.4e-12} T^{1/2} \exp\left[-\frac{12000}{T+1200}\right] \\
\gamma_{21}^{\rHt} &=& \gamma_{10}^{\rHt} 
\\
\gamma_{20}^{\rHt} &=& 0
\end{eqnarray}
where $T$ is in units of $\si{\kelvin}$, then re-scale these rates by using 
Eq.~(\ref{eq:rescale_gamma_vib}):
\begin{align}
\gamma_{\rm vib,DM}^{X}(T)\approx& \left[\racx{-1}{-5/4}{-3/4}\right]\; \gamma_{\rm vib, SM}^X(\tilde{T}_v)\,,
\end{align}
with $\tilde{T}_v\equiv \left(r_{M}^{1/2}/(r_{m}^{3/2}r_{\alpha}^2)\right)T$.
Finally, substituting into Eq.~(\ref{eq:rotn0}) along with the energy re-scaling in Eq.~(\ref{eq:rescaleEnergies}), gives the re-scaling for the cooling rate
\begin{align}
\frac{{\cal C}_{{\rm vib},n\to0,\rm DM}^X(T)}{n(\rHt)n(X)}
=
\left[\racx{}{1/4}{-5/4}\right] \;
\frac{{\cal C}_{{\rm vib},n\to0,\rm DM}^X(\tilde{T}_v)}{n(\rHt)n(X)}.
\label{eq:ldl_vib_scale}
\end{align}
At the high-density limit, collisional processes dominate so that all vibrational levels are in thermal equilibrium, and the cooling rate is given by Eq.~(\ref{eq:C_LTE}):
\begin{align}
&{\cal C}_{\rm vib, LTE}(T,n(\rHt))
\\
\nonumber
=&\,
n(\rHt)\sum_{\nu,\nu'<\nu}
g_J
\frac{\exp[-E_{\nu}/k_BT]}{{\cal Z}(T)}A_{\nu\nu'}\inc E_{\nu\nu'}\,.
\end{align}
From the re-scaling results from Section \ref{sec:basic_tools} for the Einstein $A$ coefficient [Eq.~(\ref{eq:Aquad_vib})] and the energy levels [Eq.~(\ref{eq:rescaleEnergies})], we find the 
re-scaled cooling rate in the LTE regime:
\begin{equation}
{\cal C}_{\rm{vib},\, LTE,DM}(T) 
=
\left[\racx{9}{5}{-3}\right] \;
{\cal C}_{\rm vib,\, LTE, SM}(\tilde{T}_v).
\label{eq:lte_vib_scale}
\end{equation}

\subsection{Rovibrational cooling} \label{sec:rovib_cooling_scaling}
\begin{figure*}[htbp!]
\centering
		\begin{tabular}{c c}
			\includegraphics[width=0.49\textwidth]{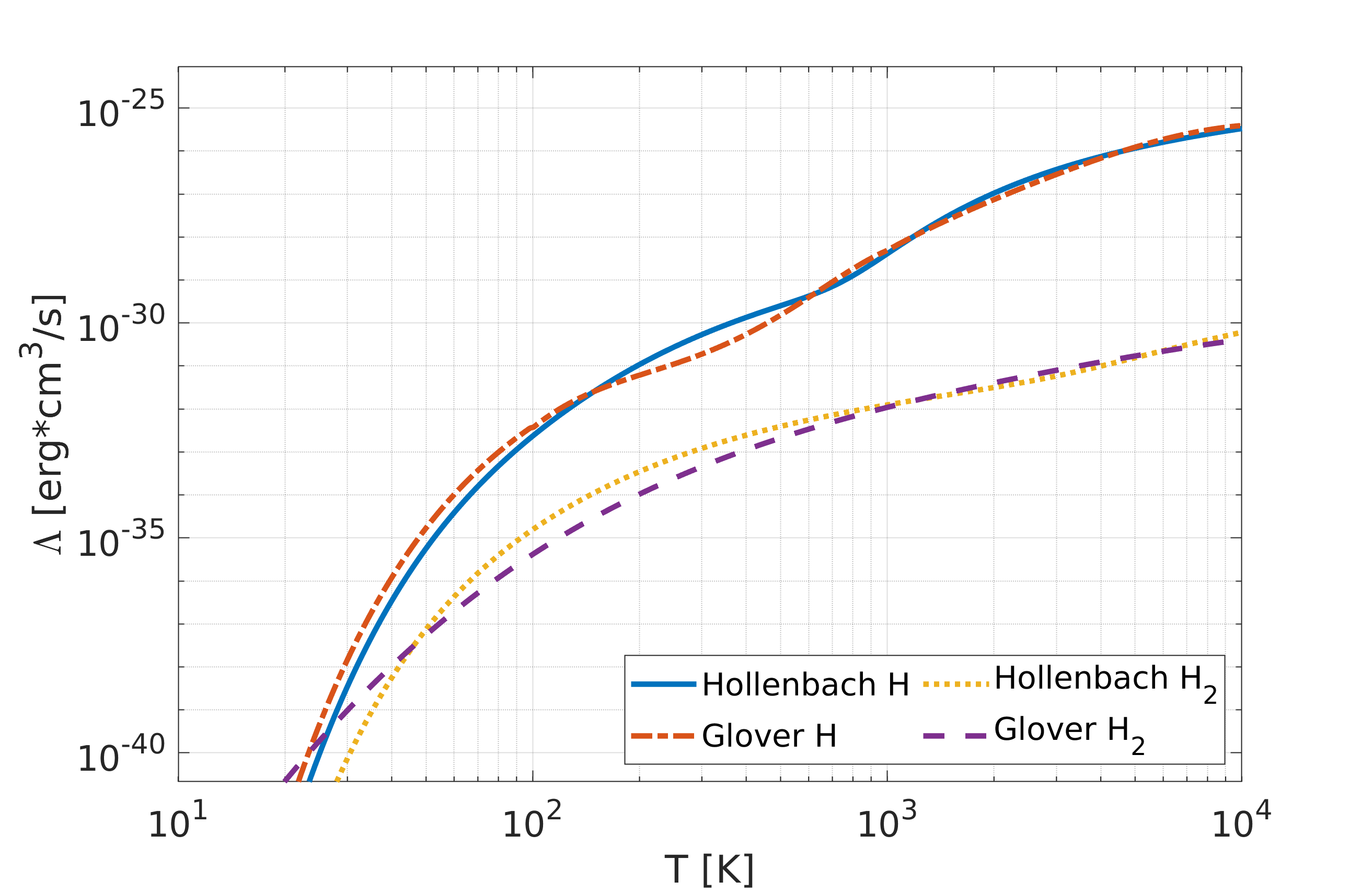} & 
			\includegraphics[width=0.49\textwidth]{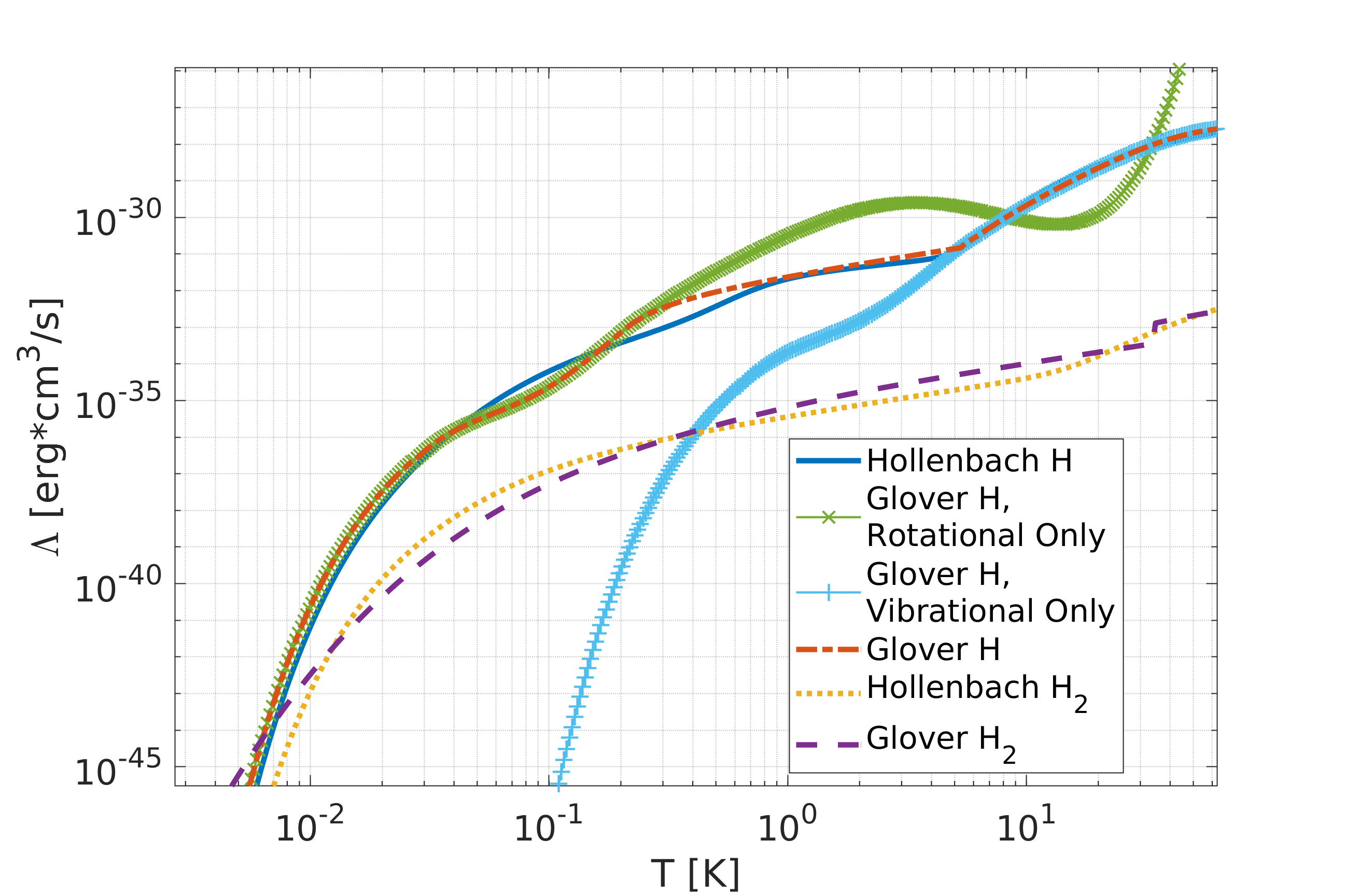} \\
			\includegraphics[width=0.49\textwidth]{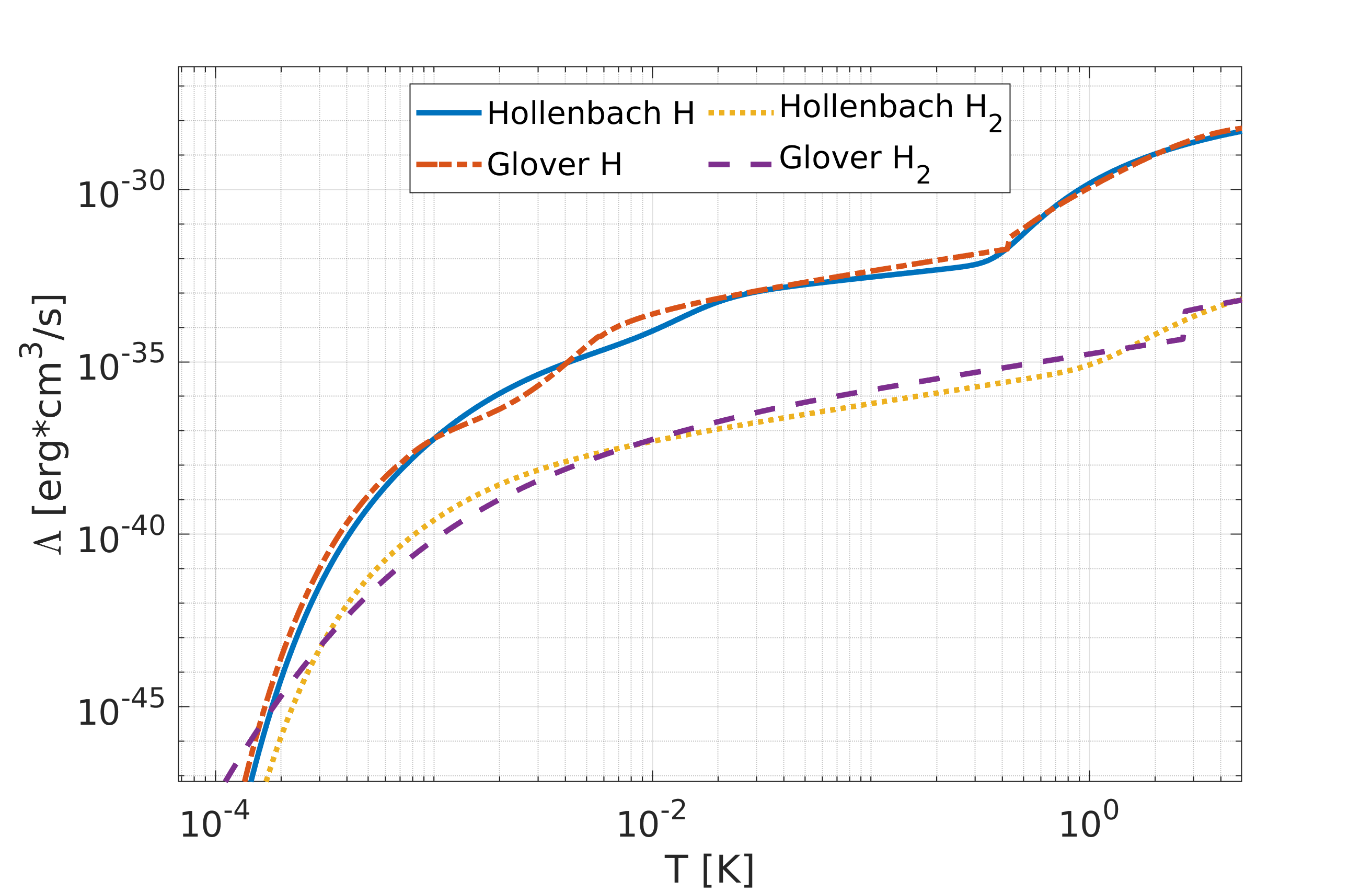} & 
			\includegraphics[width=0.49\textwidth]{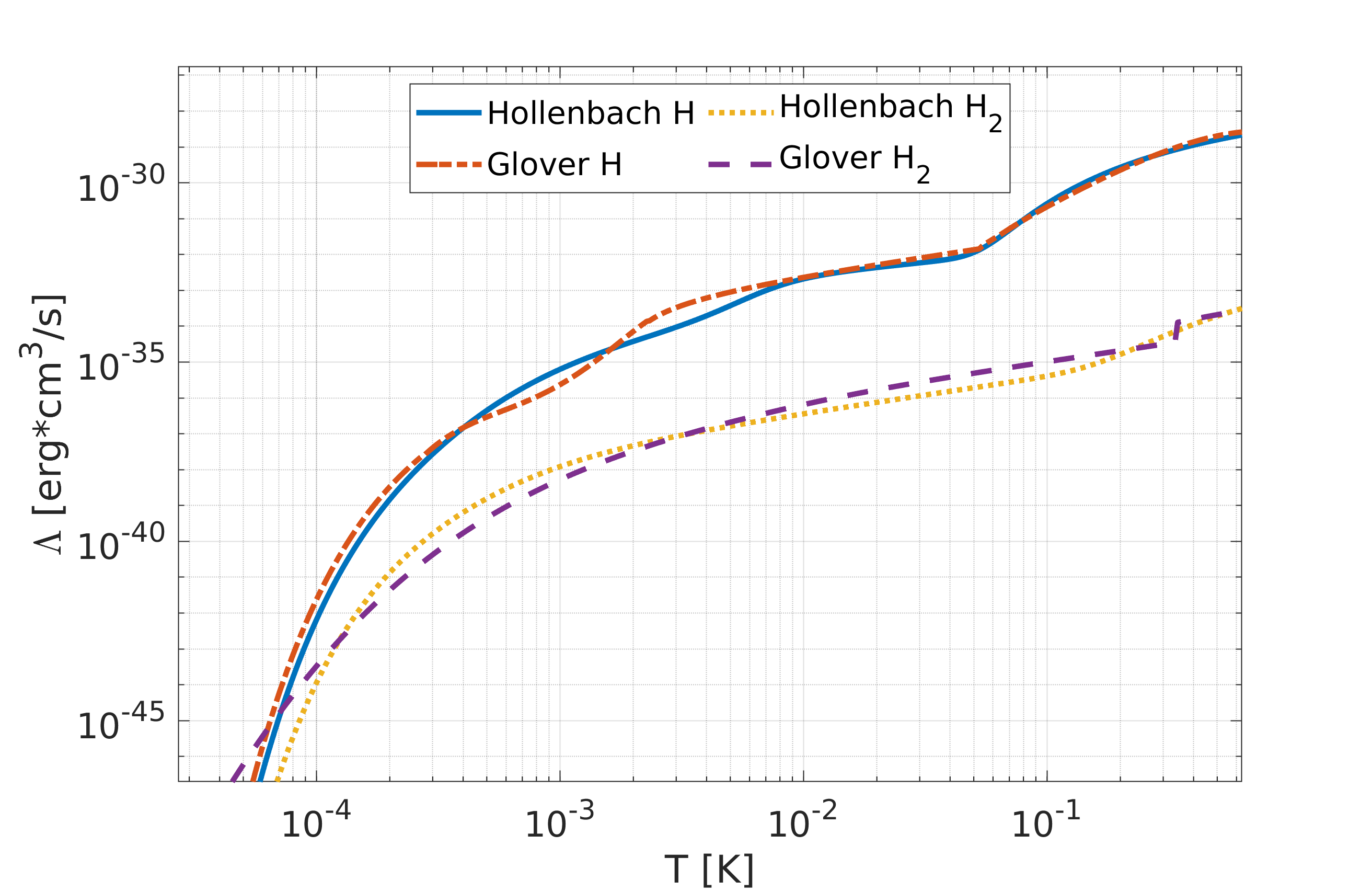} 
		\end{tabular}
		\caption{$\rHt-\{\rH,\rHt\}$ collisional cooling curves obtained from \citet{Hollenbach1979} and \citet{Glover2008} in the low-density limit   ($\Lambda_{n\rightarrow0}^{\rH,\rHt}$). All plots have $n(\rHt)=\SI{e-4}{\per\cubic\centi\meter}$ and $n(\rH)=\SI{1}{\per\cubic\centi\meter}$. The {\bf top left} panel shows a comparison of two cooling curves with Standard Model values. The {\bf top right} panel demonstrates the necessity of including both rotational and vibrational re-scaling for atomic dark matter parameters $(M,\,m,\,\alpha)=(\SI{40}{\giga\electronvolt},\,\SI{40}{\kilo\electronvolt},\,\num{0.01})$: the Glover curves are shown using rotational only (green x), vibrational only (cyan +), or both rotational and vibrational re-scaling (red dash-dot). Clearly, using only one re-scaling fails in temperature regions with a different dominant process. The lower Hollenbach and McKee curves in the upper left plot, and all curves in the lower two plots, use the combined re-scaling detailed in Appendix \ref{app:rescale_Ctot}. The ``kink" in the $\rHt$ curves arises from a combination of approximating the $\rHt$ transition temperature re-scaling and the simple gap-filling method we use. The other parameter values shown are $(M,\,m,\,\alpha)=(\SI{100}{\giga\electronvolt},\,\SI{10}{\kilo\electronvolt},\,\num{0.01})$ ({\bf bottom left}), and $(M,\,m,\,\alpha)=(\SI{40}{\giga\electronvolt},\,\SI{40}{\kilo\electronvolt},\,\num{0.001})$ ({\bf bottom right}).} 
		\label{fig:ldl_scaled_cooling}
\end{figure*}
To calculate the total cooling rate, we have to combine the rotational cooling rate and vibrational cooling rate. First of all, using the approximation in Eq.~(\ref{eq:density_interpolation}), we can combine the results of the previous sections by
\begin{align}
{\cal C}_{\rm tot}
=&
{\cal C}_{\rm rot}
+
{\cal C}_{\rm vib}
\nonumber
\\
=&
\frac{{\cal C}_{\rm rot, LTE}}{1+{\cal C}_{\rm rot, LTE}/\sum_X {\cal C}_{{\rm rot},n\to 0}^X}
\nonumber
\\
&+
\frac{{\cal C}_{\rm vib, LTE}}{1+{\cal C}_{\rm vib, LTE}/\sum_X {\cal C}_{{\rm vib},n\to 0}^X}\,,
\end{align}
with $X={\rH},~{\rH}_2$ being the colliding particles. As described in the previous sections, we use the collisional excitation rates from \citet{Hollenbach1979}.

On the other hand, most recent literature combines the rotational and vibrational cooling rates in the manner specified in \citet{Galli1998}
\begin{equation}
{\cal C}_{\rm tot} 
= \frac{{\cal C}_{\rm{rot},LTE} + {\cal C}_{\rm{vib},LTE}}{1+({\cal C}_{\rm{rot},LTE}+ {\cal C}_{\rm{vib},LTE})/{\cal C}_{{\rm tot},n\to0}}\,,
\label{eq:ga_full_sum}
\end{equation}
with 
\begin{equation}
{\cal C}_{{\rm tot},n\to0}
=
\sum_{X} \left({\cal C}_{\rm{rot},n\rightarrow 0}^{X}+{\cal C}_{\rm{vib},n\rightarrow 0}^{X}\right)\,.
\end{equation}
Since more recent rates have improved collisional coefficients \citep{Galli1998,Lee2008,Lique2015}, and include ortho-para conversions and other improvements \citep{Glover2008}, we would prefer to use the rate formulation given in Eq.~(\ref{eq:ga_full_sum}).  Our companion works \citep{Gurian2021,Ryan2021}, for example, use the rates from \citet{Glover2008,Glover2015}. 
This introduces complications, however, as the low-density limit terms ${\cal C}_{\rm{rot},n\rightarrow 0}^{\rH,\rHt}$ and ${\cal C}_{\rm{vib},n\rightarrow 0}^{\rH,\rHt}$ are usually given in the form of an analytic fit to the sum ${\cal C}_{\rm{rot},n\rightarrow 0}^{\rH,\rHt}+{\cal C}_{\rm{vib},n\rightarrow 0}^{\rH,\rHt}$, while the LTE terms are generally given individually. This poses a non-trivial problem for our purpose, because, as demonstrated in Sections \ref{sec:gen_rotation} and \ref{sec:gen_vibration}, rotational and vibrational processes are subject to different re-scalings.
In Appendix~\ref{app:rescale_Ctot}, we present a method for re-scaling the total cooling rate with the assumption that rotational cooling rate and vibrational cooling rate dominate at opposite ends of the temperature range.

In Figure \ref{fig:ldl_scaled_cooling}, we show the resulting low-density-limit total cooling rate of the hydrogenic molecule colliding with $\rH$ or $\rHt$. For literature comparison, we have plotted $\Lambda$ defined as
\begin{equation}
\Lambda^X 
= \frac{{\cal C}^X_{{\rm rot}, n\to0}+{\cal C}^X_{{\rm vib},n\to0}}
{\left(n(\rH)+n(p)+2\,n(\rHt)\right)^2}\,.
\end{equation}
For this plot, we choose the values $n(\rHt)=\SI{e-4}{\per\centi\meter\cubed}$, $n(\rH)=\SI{1}{\per\centi\meter\cubed}$, and neglect the ionized state $n(p)$. 
While simply using just the rotational or just the vibrational re-scaling is insufficient ({\it top-right} panel), the method that we outlined in Appendix~\ref{app:rescale_Ctot} provides results close to what we obtain by re-scaling the individual rotational and vibrational formulas in  \citet{Hollenbach1979} (and outlined in the previous section). This method, therefore, allows us to re-scale the modern and more accurate cooling function from \citet{Glover2008}. 
Note that the fit from \citet{Glover2008} is only well-defined in the range $\SI{10}{\kelvin}\le T \le \SI{e4}{\kelvin}$ and that this range shifts proportionally to the dark parameters. The high temperature behaviors are not a significant issue, however, as they occur above the disassociation temperature, \SI{4.48}{\electronvolt}; hence, we do not expect to have significant amounts of hydrogen molecules at that temperature.

\subsection{Collisions with other species \label{sec:other_species}}
\begin{figure*}[htbp!]
\centering
		\includegraphics[width=0.75\textwidth]{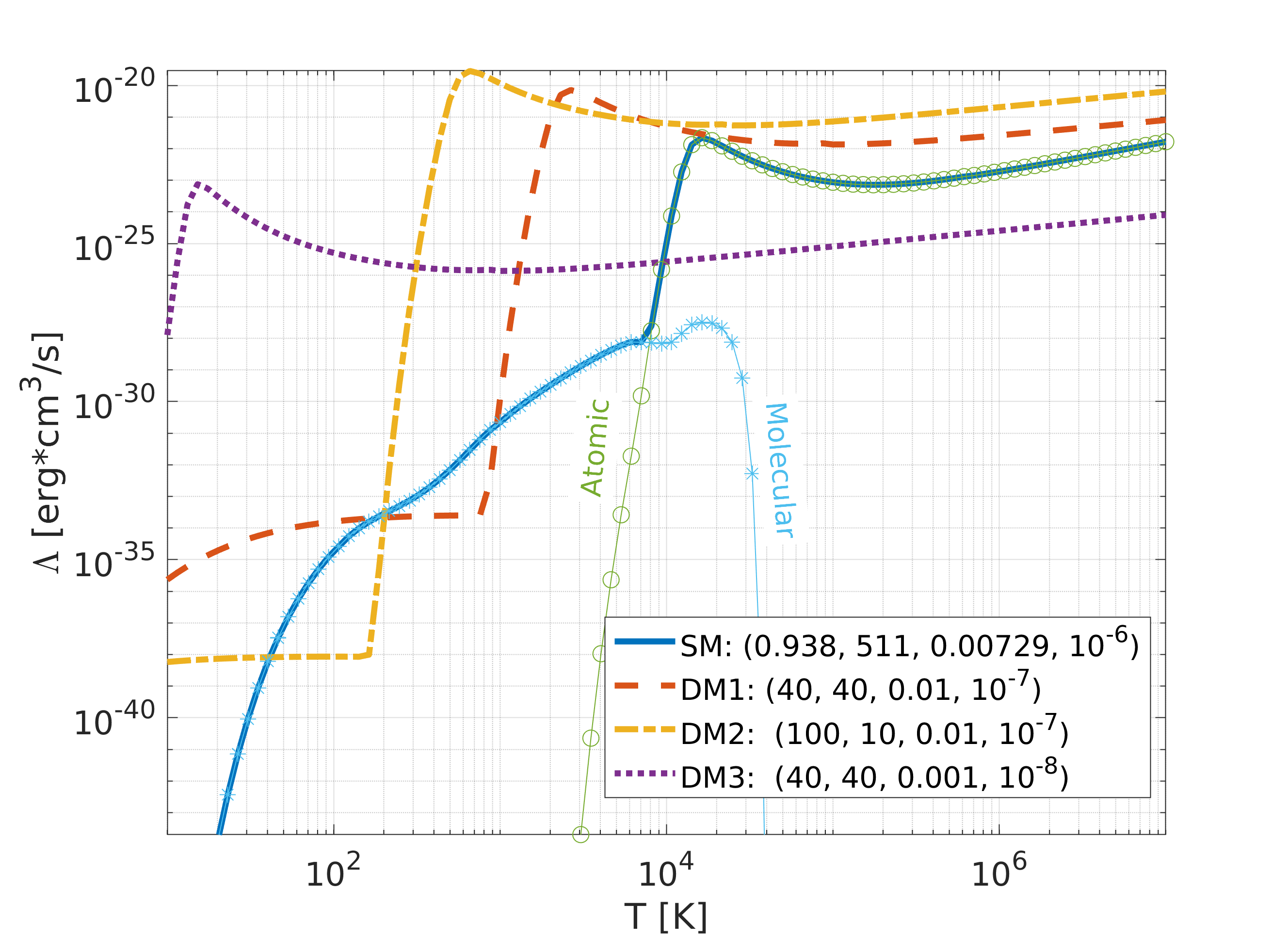} 
		\caption{Combined atomic and molecular cooling rate for atomic dark matter for the dark parameters listed, in the format $(M [\si{\giga\electronvolt}],m [\si{\kilo\electronvolt}],\alpha,x_{\rHt})$. The atomic cooling rates were derived in \citet{Rosenberg2017} and the $\rHt$ fraction, $x_{\rHt}$, represents the cosmological fraction found in \citet{Gurian2021}. For the Standard Model case, we have highlighted the atomic (green circles) and molecular (cyan stars) contributions. Note that the molecular cooling can be significantly reduced compared to atomic cooling as the parameters are varied from the Standard Model values.}
		\label{fig:total_cooling_curve}
\end{figure*}
Atomic and molecular hydrogen are not the only collisional partners in the Standard Model primordial gas. \citet{Glover2008} showed that collisions with He, protons, and electrons also contribute to cooling, although with slower rates than the hydrogenic cooling due to their lower abundances. Even at low abundances, however, the proton collisions can be important at low temperatures, both as a direct additional source of cooling and through an effect on the ortho-para ratio. The atomic dark matter model has no dark neutrons and no dark helium. This section, therefore, focuses on re-scaling the cooling rate involving collisions with protons and electrons.

The $\rHt-p$ scattering rate re-scaling for a rotational transition and the effect of proton collisions on the ortho-para ratio are covered in more detail in Appendix \ref{app:orthopara}, where we assume the cross-section can be approximated by a Langevin rate (Section \ref{sec:detailed_balance_example} for more details). The re-scaling for a vibrational transition follows from using the vibrational energy re-scaling in Eq.~\ref{eq:op_p_rate_scale}, giving 
$\gamma_{{\rm DM},v\rightarrow v'}(T) = \left[\racx{-1}{-3/2}{-1/2}\right]\gamma_{{\rm SM},v\rightarrow v'}(\tilde{T}_v)$. The cooling rates are then derived from the appropriate substitutions into Eqs.~(\ref{eq:rotn0}) and (\ref{eq:vibn0}):
\begin{align}
    \frac{{\cal C}_{{\rm rot},n\to 0,\rm DM}^p(T)}{n(\rHt)n(p)} &=
\left[\racx{}{1/2}{-3/2}\right] \; \frac{{\cal C}_{{\rm rot},n\to 0,\rm SM}^p(\tilde{T}_r)}{n(\rHt)n(p)} \label{eq:proton_rot_scaling}\\
\frac{{\cal C}_{{\rm vib},n\to 0,\rm DM}^p(T)}{n(\rHt)n(p)} &=
\left[\rax{}{-1}\right] \; \frac{{\cal C}_{{\rm vib},n\to 0,\rm SM}^p(\tilde{T}_v)}{n(\rHt)n(p)}
.\label{eq:proton_vib_scaling}
\end{align}

We also use the Langevin-type cross section for the $\rHt-e$ collision, which leads to a rate re-scaling of $\gamma_{{\rm DM},j\rightarrow j'}(T) = \left[\rac{-1}{-2}\right]\gamma_{{\rm SM},j\rightarrow j'}(\tilde{T}_r)$ and $\gamma_{{\rm DM},v\rightarrow v'}(T) = \left[\rac{-1}{-2}\right]\gamma_{{\rm SM},v\rightarrow v'}(\tilde{T}_v)$. Again, substituting appropriately into Eqs.~(\ref{eq:rotn0}) and (\ref{eq:vibn0}) we have
\begin{align}
    \frac{{\cal C}_{{\rm rot},n\to0,\rm DM}^e(T)}{n(\rHt)n(e)} &=
\left[\rax{}{-1}\right] \; \frac{{\cal C}_{{\rm rot},n\to0,\rm SM}^e(\tilde{T}_r)}{n(\rHt)n(e)} \label{eq:electron_rot_scaling}\\
\frac{{\cal C}_{{\rm vib},n\to0,\rm DM}^e(T)}{n(\rHt)n(p)} &=
\left[\racx{}{-1/2}{-1/2}\right] \; \frac{{\cal C}_{{\rm vib},n\to0,\rm SM}^e(\tilde{T}_v)}{n(\rHt)n(e)}
.\label{eq:electron_vib_scaling}
\end{align}

Like the $\rHt-\{\rH,\rHt\}$ collisions, the rotational and vibrational cooling rates for both the $\rHt-p$ and $\rHt-e$ collisions have different re-scalings, yet the Standard Model rates are fits to the sum. We would prefer to account for this using the approach in Appendix \ref{app:rescale_Ctot}, however, we have been unable to obtain a clear separation between the two temperature regimes in the Standard Model rate. As such, we have used the rotational scaling [Eqs.~\ref{eq:proton_rot_scaling} and \ref{eq:electron_rot_scaling}] for the entirety of the temperature range. This leads to cooling that is stronger by a factor of $\sqrt{\rc{}/\rx{}}$ in the vibrational temperature regime.  We believe this an  acceptable approximation for two reasons: first, the rates have the largest relative contribution in the low temperature regime \citep{Glover2008} where the vibrational portion is unlikely to be relevant, and more importantly, we expect the overall contribution from $\rHt-\{e,p\}$ cooling to be small due to relative abundances for the majority of cases. For example, for the parameter values used in the top right panel of Figure \ref{fig:ldl_scaled_cooling}, the $\rHt-\{e,p\}$ collisional cooling does not contribute meaningfully, regardless of the re-scaling used, until $\sim$\SI{1300}{\kelvin}, where much stronger atomic processes become relevant.

\subsection{Summary of cooling results} \label{sec:cool_summary}
Finally, in Figure \ref{fig:total_cooling_curve}, we show the entire atomic-dark-matter cooling rate $\Lambda$, including both the molecular rovibrational cooling defined above and the combination of cooling processes provided in \citet{Rosenberg2017}. We have computed the rate for the same dark parameter sets used in Figure \ref{fig:ldl_scaled_cooling} to demonstrate the significant behavioral changes induced by varying the parameters. We have also varied the ionization fraction, assuming chemical equilibrium, and the dark-atomic hydrogen fractions $x_{\rHt}$ to match the cosmological abundance obtained in \citet{Gurian2021}.

\section{Application: Reaction rates} \label{sec:reactionrates}
The molecular line cooling in the previous section is proportional to the abundance of $\rH$ and $\rHt$. In order to complete the molecular line cooling calculation for the dark-atomic model, therefore, we have to determine their abundances. These abundances are generally computed numerically from a set of coupled differential rate equations similar to Eq.~(\ref{eq:two_state_density_differential}), for example, 
\begin{equation}
\frac{dn_{\rHt}}{dt} 
= \sum_{j\in F_i}\left(\gamma_j\prod_{k\in R_j}n_k\right) - \sum_{j\in D_i}\left(\gamma_j\prod_{k\in R_j}n_k\right),
\end{equation}
where the first sum represents the source term including all chemical reactions that contain $\rHt$ as one of the products (Formation reactions), and the second sum represents the sink term including all chemical reactions that contain $\rHt$ as one of the reactants (Destruction reactions). Thus, to determine the species' abundance evolution in the atomic-dark-matter model, we need to know how to re-scale the reaction rates ($\gamma$) to give the leading dependence on the dark-matter parameters. 

We shall present the re-scaling of these rates in this section. In the first sub-section, we give an overview of how the re-scaling procedure works and present a summary table of our results. In the remaining sub-sections we provide the details.
	
\subsection{Overview of method and summary of result}\label{sec:rate_overview_summary}
Just like the case for the collisional excitation rates that we studied in Section~\ref{sec:scattering}, from a given standard-model rate $\gamma_{\rm SM}$, the corresponding rate for the dark-matter model can be estimated by an overall re-scaling factor multiplying $\gamma_{\rm SM}$ evaluated at a re-scaled temperature $\tilde{T}$. That is,
\begin{align} 
\label{eq:overall_rescaled_rate}
\gamma_{\rm DM}(T) 
=
g(\ra{},\rc{},\rx{}) \; \gamma_{\rm SM}\left(\tilde{T}_{\inc}\right)\,,
\end{align}
where $g(\ra{},\rc{},\rx{})$ is the overall re-scaling, and 
$\tilde{T}_{\inc}=T/\rde$ with $\rde$ the re-scaling of the primary binding energy relevant for the reaction. 
	
For non-photochemistry reactions, the calculations of interaction rates are similar to Eq.~(\ref{eq:approxHH2}) from Section \ref{sec:basic_tools}, but with the added possibility of chemical reactions instead of just pure scattering. Assuming that all particles in the gas are at the same temperature, the reaction rate is
\begin{equation} 
	\gamma_{\rm DM, non-\gamma} 
	= \langle\sigma v\rangle
	\propto\sqrt{\frac{T}{\mu}}\int_0^{\infty}\sigma(x;y) x^3 e^{-x^2} \integrand{x}.
		\label{eq:reaction_rate_basic}
\end{equation}
Here, we reuse the two dimensionless parameters defined in Section \ref{sec:basic_tools}:
\begin{eqnarray}
\label{eq:dimlessx}
x^2 &=& \frac{\KE}{k_B\,T} = \frac{\mu\,v^2}{2\,k_B\,T}\,,
\\
\label{eq:dimlessy}
y^2 &=& \frac{\inc E}{k_B\,T}\,,
\end{eqnarray}
where $\Delta E$ is the relevant binding energy of the chemical reactions. Note that, for chemical reactions, we cannot assume that the cross-section scales only as some effective size of the particles. Instead, we need to determine in more detail how the cross section depends on kinetic energy and binding energies, and how strongly the reaction depends on $\alpha$. 
	
\begin{table*}
\begin{rotatetable*}
		\scriptsize
		\centering
		\begin{threeparttable}
		\begin{tabular}{l l c c c c c}
			\toprule 
			\multirow{2}{*}{\#} & \multirow{2}{*}{Reaction} & \multirow{2}{*}{Cross section source}  & \multirow{2}{*}{$\sigma$} & Re-scaling pre-factor & \multirow{2}{*}{b} &   \multirow{2}{*}{Additional notes} \\
			& & & & $g(\ra{},\rc{},\rx{})$ &  & \\
			\midrule
			1 & $p+e\rightarrow \rH + \gamma$(\ref{sec:Hrecomb}) & \citet{Mo2010} & $\frac{\alpha^5}{\KE(\KE+\inc E)} $  & $\rac{2}{-2}$  & $-0.62,-1.15$ &  \tnote{a} \tnote{,} \tnote{b} \\ 
			2 & $\rH + \gamma\rightarrow p+e$(\ref{sec:Hrecomb}) & \citet{Mo2010} & $\mu\, \alpha^5\frac{1}{(\KE+\inc E)^3}$ & $\rac{5}{}$ & 0.88, 0.35 & \tnote{c}\\
			3 & $\rH+e \rightarrow \rH^- + \gamma$ (\ref{sec:h_e_detach})  & \citet{deJong1972}  & $\frac{ \alpha}{\mu^2}\; \frac{\inc E^{1/2}\, \KE^{1/2}}{(\KE+\inc E)} $ & $\rac{2}{-2}$  & 0.928  & \tnote{a} \tnote{,} \tnote{d} \\ 
			4 & $\rH^- + \gamma \rightarrow \rH + e$ & \citet{Armstrong1963} & $ \frac{\alpha}{\mu}\frac{\inc E^{1/2} K^{3/2}}{(\KE+\inc E)^3}$ & $\rac{5}{}$  &  2.13 & \tnote{c} \tnote{,}  \tnote{d} \\
			5 & $\rH^- + \rH \rightarrow \rHt+ e$ & \citet{Browne1969} & $ \sqrt{\frac{\alpha \, a_0^3}{\KE}} $ & $\racx{-1}{-3/2}{-1/2}$  & 0 & \tnote{e} \tnote{,} \tnote{f} \tnote{,} \tnote{g} \\
			7 & $\rH^- + p \rightarrow 2 \rH$ (\ref{sec:mutual neutralization})  & \citet{Bates1955}  & $\alpha\,a_0^2 \sqrt{\mu} \, \frac{\sqrt{\KE+\inc E} }{\KE \, \inc E}$ &  $\rac{-3}{-3}$ & $-\frac{1}{2}$ & \tnote{e} \tnote{,} \tnote{h} \\
			8 & $\rH + p \rightarrow \rHt^+ + \gamma$ & \citet{Stancil1993} & $\frac{({\KE} +\inc E)^3 \alpha^{4}}{\eh^{3}{\KE}^{3/2} M^{1/2}}$ & $\racx{2}{-1}{-1}$  & 1.8 & \tnote{i}\\
			9 & $\rHt^+ + \gamma \rightarrow \rH + p $  & \citet{Stancil1993} & $\left(\frac{\mu \, v}{h\, \nu}\right)^2\,\frac{({\KE} +\inc E)^3 \alpha^{4}}{\eh^{3}{\KE}^{3/2} M^{1/2}}$ & $\racx{5}{1/2}{1/2}$  & 1.59 & \tnote{c} \tnote{,} \tnote{j}\\
			10 & $\rHt^+ + \rH \rightarrow \rHt+p$(\ref{sec:detailed_balance_example}) &\citet{Galli1998}  & $\sqrt{\frac{\alpha \, a_0^3}{\KE}} $ & $\racx{-1}{-3/2}{-1/2}$  & 0 & \tnote{g}\\
			13 & $\rHt^+ + \rHt \rightarrow \rH_{3}^+ +\rH$ &---\textquotedbl--- & ---\textquotedbl--- & ---\textquotedbl---   & 0 & \tnote{g}\\

			15 & $\rHt + p \rightarrow \rHt^+ + \rH$(\ref{sec:detailed_balance_example}) & ---\textquotedbl--- & ---\textquotedbl--- & ---\textquotedbl---    & 0 &  \tnote{g} \tnote{,} \tnote{j}\\
			20 & $ \rH_{3}^+ + e\rightarrow \rH +\rHt $ &\citet{Draine2011} & $\frac{\alpha \, a_0}{\KE}$ & $\racx{-1}{-2}{}$  & -0.65 & \tnote{k}\\
			* & $\rHt + \rH \rightarrow 3 \rH$ (\ref{sec:h2_diss}) & Hard Sphere & $a_0^2$ & $\racx{-1}{-3/2}{-1/2}$ & 0 & \tnote{l} \\
			3B1 & $3\rH \rightarrow \rHt+\rH$ (\ref{sec:Three-Body}) & Hard Sphere/Detailed Balance & $a_0^2\left[\frac{n_{\rHt}}{n^2_\rH}\right]_{\rm LTE}$ & $\racx{-4}{-4}{-1}$ & -1 & \tnote{j} \tnote{,} \tnote{m} \\
			3B2 & $\rHt+2\rH \rightarrow 2\rHt$ (\ref{sec:Three-Body}) & ---\textquotedbl--- & ---\textquotedbl--- & ---\textquotedbl---  & -1 & \tnote{j} \tnote{,} \tnote{m} \\
			3B3 & $2\rH + \rH^+ \rightarrow \rHt+\rH^+$ (\ref{sec:Three-Body}) &---\textquotedbl---& ---\textquotedbl--- &---\textquotedbl---  & -1& \tnote{j} \tnote{,} \tnote{m} \\
			3B4 & $2\rH + \rH^+ \rightarrow \rHt^+ + \rH$ (\ref{sec:Three-Body}) & ---\textquotedbl---& ---\textquotedbl---& ---\textquotedbl---  & -1 & \tnote{j} \tnote{,} \tnote{m} \\
			\bottomrule
		\end{tabular}
\caption{Table of Reactions. Reactions are numbered according to \citet{Galli1998} and the section numbers in parentheses indicate where details can be found in the text. Only the parametric dependence on key quantities are shown for the cross sections (no numerical factors). The reactions included in this table are only those that were considered ``important" in \citet{Galli1998},  are required for our companion paper \citep{Gurian2021}, or are three-body reactions.  To leading order the Standard Model rates can be approximated as $\gamma \propto g(\ra{},\rc{},\rx{}) \; (T/\rde)^b$ and values for $b$ are from Galli and Palla unless otherwise noted. All reactions listed have binding energy proportional to $\eh$, so here $\rde=\rac{2}{}$. For reactions 1-4 and 20, the reduced mass, $\mu$, is proportional to the dark electron mass, $m$. For the remaining reactions, $\mu$ is proportional to the dark proton mass, $M$.
\label{tab:reactions_table_part1}}
		\begin{tablenotes}
			\item [a] Milne Relation
			\item [b] The $b$ values are an expansion of the Case B recombination coefficient in low and high temperature regimes (respectively), from \citet{Pequignot1991}. The re-scaling of the Case A coefficient is identical, although $b$ differs. 
			\item [c] Photoionization re-scaling
			\item [d] Effective Range approximation
			\item [e] These rates are uncertain by up to an order of magnitude \citet{Glover2008}
			\item [f] Rate is constant for $T\le\SI{300}{\kelvin}$, and very uncertain for $T>\SI{300}{\kelvin}$ \citet{Galli1998}
			\item [g] Langevin Reaction
			\item [h] Landau-Zener method
			\item [i] $b$ value from \citet{Stancil1998}
			\item [j] General Detailed Balance
			\item [k] Coulomb focusing
			\item [l] The $b$ value is from \citet{Lepp1983} and does not account for density dependence.
			\item [m] The $b$ value is from \citet{Yoshida_2006}.
		\end{tablenotes}
		\end{threeparttable}
\end{rotatetable*}
\end{table*}
	
Even for chemical reactions, however, we can organize the scattering rate, Eq.~(\ref{eq:reaction_rate_basic}), as a product of the universal thermal factor, an effective ``impact parameter" $\tilde{R}$ (with dimensions of length) squared, and a dimensionless integral, analogously to Eq.~(\ref{eq:approxHH2}). The reaction-dependent part of the pre-factor, $g(\ra{},\rc{},\rx{})$, then depends on the re-scaling of the ``impact parameter". In general, this will be an algebraic combination of the masses, $\alpha$, and $T$ (we do not need to track whether the $T$ comes from $x$ or $y$ substitutions). Considering only the dominant parametric dependence as power laws, the pre-factor can be written
\begin{equation}
g(\ra{},\rc{},\rx{}) = \left(r_{\mu}^{-1/2} \rde^{1/2} \right) \left(\racx{R_{\alpha}}{R_{m}}{R_M} \rde^{R_T}\right),
\label{eq:overall_scaling_factor}
\end{equation}
where the terms in the first parenthesis come from the universal thermal factor, and the terms in the second are from the parametric dependencies of $\tilde{R}$. 
	
For photochemistry reactions, we have different factors and integrals, but the overall approach is identical, with the substitution of the radiation temperature, $T_{\gamma}$, instead of the particle temperature, $T$, and a slightly different universal thermal factor. Thus, we have
\begin{align} 
\gamma_{\rm DM, \gamma}(T_{\gamma}) 
&=
2c\int_{\Delta E/h}^\infty
d\nu
\frac{\nu^2 \sigma_{\rm photo}(\nu)}{e^{h\nu/k_BT_{\gamma}}- 1}
\nonumber\\
&= 
g(\ra{},\rc{},\rx{}) \; \gamma_{\rm SM, \gamma}\left(\frac{T_{\gamma}}{\rde}\right),
\end{align}
with
\begin{equation}
g(\ra{},\rc{},\rx{}) = \left( \rde^{3} \right) \left(\racx{R_{\alpha}}{R_{m}}{R_M} \rde^{R_T}\right).
\label{eq:overall_scaling_factor_photo}
\end{equation}

Table \ref{tab:reactions_table_part1} summarizes the main results of this section. This table contains  ``the most important'' molecular hydrogen reactions as designated by \citet{Galli1998} (see also \citet{Abel1997}),  as well as several additional reactions needed for accurate results in our companion papers  \citep{Gurian2021,Ryan2021}.
The table shows the parametric dependence of the cross sections, the re-scaling pre-factor (capturing only the dominant parametric dependence), and the dominant power-law temperature dependence of the rate as $b$.
Standard Model rates frequently have a temperature dependence that can be captured to a first approximation by a power law $\gamma_{\rm SM}\propto T^b$, and it is often useful to see this power law in order to understand the primary parametric dependence of a reaction. 
However, the full temperature dependence known for the Standard Model can of course be used (as long as the re-scaled temperature is used throughout) and is recommended.
In the following sub-sections we go through detailed derivations of a few representative cases.	
\paragraph{Principle of detailed balance}
\label{sec:forward_backward}
Several reactions in Table \ref{tab:reactions_table_part1} are inverses of each other, and we can exploit that property to compute some re-scalings using the principal of detailed balance \citep{Mo2010}. In thermal equilibrium, forward and backward reaction rates must match, leading to the following relationship between their cross sections: 
	\begin{equation}
		\sigma_f \propto \left(\frac{g_b}{g_f}\right)\,\left(\frac{p_b}{p_f}\right)^2 \, \sigma_b,
		\label{eq:detailed_balance}
	\end{equation}
	
While $p_b$ and $p_f$ are reaction specific, recombination/photoionization-type reactions all have $p_b=(h\,\nu)/c$ and $p_f=m_{\rm ej} \,v$ and we obtain the Milne Relation \citep{Mo2010}, 
\begin{equation}
\sigma_{\rm rec}(v,n) = \frac{g_n}{g_{n+1}} \left(\frac{h \, \nu}{m_{\rm ej} \,v\, c }\right)^2  \sigma_{\rm pi}(\nu,n),
\label{eq:Milne}
\end{equation}
where $\sigma_{\rm rec}$ is the recombination cross section, $\sigma_{\rm pi}$ is the photonionization cross section, $h\,\nu$ is the photon energy, and $m_{\rm ej}\, v$ is the momentum of the ejected particle (e.g. an electron in standard hydrogen recombination; see Section~\ref{sec:Hrecomb}). Generally, we have $h\, \nu=\KE+\inc E=k_B\,T\left(x^2+y^2\right)$ through conservation of energy (with $x$ and $y$ defined in Eqs.~(\ref{eq:dimlessx}) - (\ref{eq:dimlessy})) and usually $m_{\rm ej}\approx\mu$, such that 
\begin{equation}
\sigma_{\rm rec} \propto 
\frac{ T}{\mu}\frac{\left(x^2+y^2\right)^2}{ x^2} \sigma_{\rm pi}
\end{equation}

\subsection{Reaction 1: Hydrogen recombination}
\label{sec:Hrecomb}
Let us start with one of the simplest reactions, hydrogen recombination: $p+e\rightarrow {\rH}+\gamma$. This reaction is often divided into Case A, where the hydrogen atoms are sparse enough to recombine directly to the ground state, and Case B, where the hydrogen atoms are too dense for the Lyman continuum photons to escape, restricting recombination to excited states. The dependence on dark-sector parameters has been worked out already in \citet{Rosenberg2017} (Case A) and in \citet{Hart_2017} (Case B), and so this is a good example to demonstrate the validity of the re-scaling procedure in this work. The cross section scales in the same way for Case A and Case B, but the reaction rates of course differ. In Table \ref{tab:reactions_table_part1}, we have used $b$ values derived from the Case B rate of \citet{Pequignot1991} in order to have a relevant comparison to the photoionization rate (in Case A, the appropriate comparison would be with the collisional ionization rate).  In this section, we obtain the expression for the recombination cross-section using the principal of detailed balance and the photoionization cross section, and apply this expression to the Case A rate of \citet{Cen1992}.

The photoionization cross section from the ground state of a hydrogenic atom is \citep{Mo2010} 
\begin{equation}
\sigma_{\rm pi}(\nu) = 
\frac{g_{\rm bf} }{Z^2} 
\frac{64 \pi}{3\sqrt{3}}\frac{h^2}{\alpha m^2 c^2}\left[\frac{E_{\rH}}{h\, \nu}\right]^3 
\propto  \alpha \, a_0^2 \, \left(\frac{E_{\rH}}{h \, \nu}\right)^3,
\end{equation}
where $E_{\rH}=\alpha/(2 a_0)$ is the binding energy of hydrogen, $g_{\rm bf}$ is the bound-free Gaunt factor, and $Z$ is the nuclear charge ($Z=1$ in the atomic dark matter case).
From conservation of energy, $h\,\nu=\KE+\inc E$, and we have 
\begin{equation}
\sigma_{\rm pi} \propto \alpha a_0^2 \frac{\left(\frac{\alpha}{a_0}\right)^3}{(\KE+\inc E)^3} = m\,\alpha^5\frac{1}{(k_B \,T(x^2+y^2))^3}.
\end{equation}
Now that we have the photoionization cross section, application of the Milne relation [Eq.~(\ref{eq:Milne})] with $\mu=m$ gives
\begin{equation}
\sigma_{\rm rec} \propto \left(\frac{\alpha^5}{T^2}\right)\,\left(\frac{1}{x^2\,(x^2+y^2)}\right).
\end{equation}
Substituting this into Eq.~(\ref{eq:overall_scaling_factor}) and defining 
$\tilde{T}_a = T/(\rac{2}{})$
as the atomic energy analog of $\tilde{T}_r$ and $\tilde{T}_v$, we obtain our final re-scaling
\begin{equation}
\gamma_{\rm rec,DM}(T) = \left[\rac{2}{-2}\right]\; \gamma_{\rm rec,SM}(\tilde{T}_a).
\end{equation}
which matches the results in \citet{Rosenberg2017} and \citet{Hart_2017}. 

\begin{figure*}[htbp!]
\centering
		\begin{tabular}{c c}
			\includegraphics[width=0.49\textwidth]{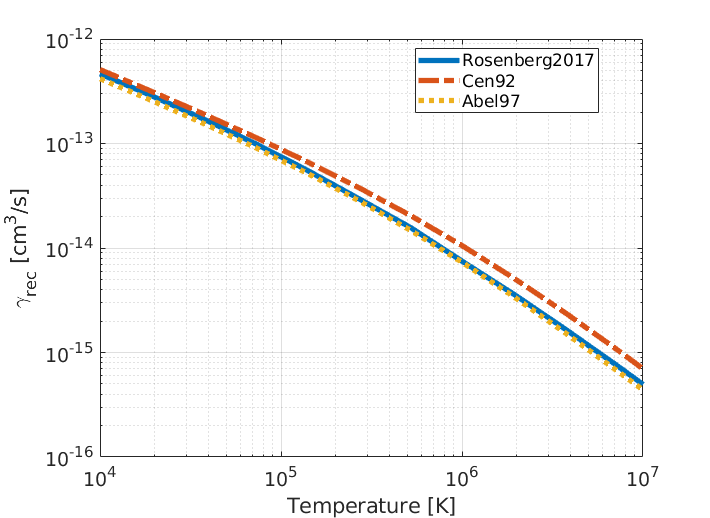} & 
			\includegraphics[width=0.49\textwidth]{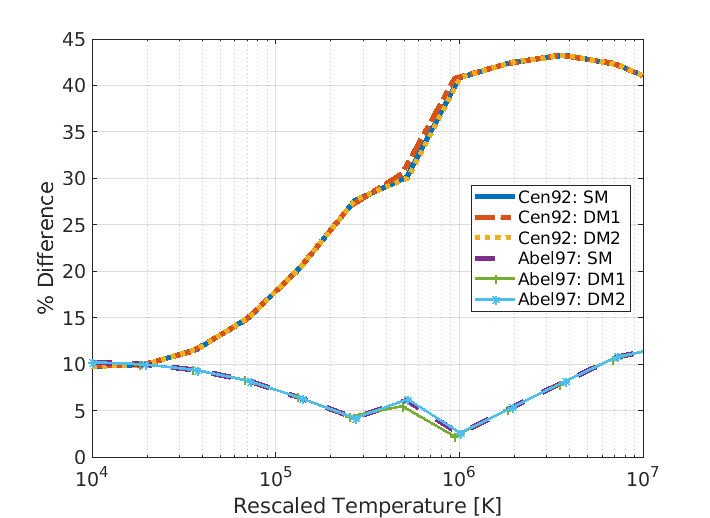} 
		\end{tabular} 
		\caption{ Comparisons between our re-scaled recombination rates from Eq.~(\ref{eq:dark_hydrogen_rescale_recomb_rate}) and (\ref{eq:abel_recombination_rate}) based on \citet{Cen1992} and \citet{Abel1997} and the full rate given in \citet{Rosenberg2017}. On the left we plot the Standard Model rates, equivalent to setting $\ra{}=\rc{}=\rx{}=1$. On the right we compute the percent difference between the Rosenberg and Cen and Rosenberg and Abel curves, i.e. $\text{\% Difference}(\rm{Ros},\rm{Cen})=100\% \frac{\left|\gamma_{\rm Ros}-\gamma_{\rm Cen}\right|}{\gamma_{\rm Ros}}$. We have plotted the rates using  $m=\{511,40,10\}\si{\kilo\electronvolt}$, $M=\{0.938,40,100\}\si{\giga\electronvolt}$, and $\alpha=\{\num{0.00729},\num{0.01},\num{0.01}\}$, labeled \{SM,DM1,DM2\}, as a function of the re-scaled temperature, $\tilde{T}_a=\rac{-2}{-1}\,T$.  Note that the increased difference between the Rosenberg and Cen curves in both plots arises from Cen being a simplified analytic fit. 
		\label{fig:recomb_comparison}
		}
	\end{figure*}

\begin{widetext}	
A simplified Standard Model recombination rate is given by \citet{Cen1992,Mo2010}:
\begin{align}
\gamma_{\rm rec,SM}(T) &= \SI{8.40e-11}{\centi\meter\cubed\per\second}T^{-1/2}\left(\frac{T}{\SI{e3}{\kelvin}}\right)^{-0.2}
\left(1+\left(\frac{T}{\SI{e6}{\kelvin}}\right)^{0.7}\right)^{-1},
\end{align}
and we find the corresponding dark recombination rate following the procedure in Section~\ref{sec:rate_overview_summary} as
\begin{align}
\gamma_{\rm rec,DM}(T) &= \SI{2.66e-12}{\centi\meter\cubed\per\second}\left[\rac{2}{-2}\right]\left(\frac{T/(\rac{2}{})}{\SI{e3}{\kelvin} }\right)^{-0.7} \left(1 + \left(\frac{T/(\rac{2}{})}{ \SI{e6}{\kelvin}} \right)^{0.7}\right)^{-1} \nonumber
\\
&\approx
\begin{dcases}
	\begin{rcases}
	    \num{2.66e-12}\left[\rac{3.4}{-1.3}\right] \left(\frac{T}{\SI{e3}{\kelvin}}\right)^{-0.7} & T\ll\rac{2}{}\SI{e6}{\kelvin} \\
		\num{2.11e-14}\left[\rac{4.8}{-0.6}\right] \left(\frac{T}{\SI{e6}{\kelvin}}\right)^{-1.4} & T\gg\rac{2}{}\SI{e6}{\kelvin}
\label{eq:dark_hydrogen_rescale_recomb_rate}
	\end{rcases}
\end{dcases}\si{\centi\meter\cubed\per\second}
\end{align}
\end{widetext}
In Figure \ref{fig:recomb_comparison}, we show that this re-scaling provides a good approximation of the true dark-recombination rate and that we can improve on these results simply by using the same re-scaling but a more accurate Standard Model rate with the form,
\begin{align}
\gamma_{\rm rec,SM}(T) = \exp\left(\sum_{n=0}^{9} A_n \log(T)^n \right),
\label{eq:abel_recombination_rate}
\end{align}
where the coefficients are given in \citet{Abel1997}.
We compare the re-scaled recombination rates, Eq.~(\ref{eq:dark_hydrogen_rescale_recomb_rate}) and the re-scaled version of Eq.~(\ref{eq:abel_recombination_rate}) with the first-principle calculation from \citet{Rosenberg2017} in Fig.~\ref{fig:recomb_comparison}.

\subsection{Reaction 3 and 4: Hydrogen electron attachment and $\rH^-$ photodetachment} \label{sec:h_e_detach}
The hydrogen electron attachment reaction, $\rH+e^- \rightarrow {\rH}^- + \gamma$, can be effectively treated as a screened proton capturing a free electron, in a similar process to standard hydrogen recombination \citep{Ohmura1960}. The reaction rate can be computed in analogous fashion, but where the $\rH^-$ photodetachment cross section is used instead of the $\rH$ photoionization cross section. Thus, we start with the photodetachment cross section, where we use results from effective range theory \citep{Armstrong1963,Ohmura1960} and conservation of energy ($h\, \nu=\inc E +\KE = \inc E + p_e^2/(2m)$, where $\inc E$ is simply the $\rH^-$ ground state energy, proportional to $\eh$) such that 
\begin{align}
\sigma_{\rm pd}(\nu) &= \frac{1}{1-\gamma\varrho}\frac{16 \pi}{3}\frac{\alpha \hbar^2}{\mu}\frac{\inc E^{1/2} (h\nu-\inc E)^{3/2}}{\left(p_e^2/2m+\inc E\right)^3} \nonumber \\
&\propto \frac{1}{1-\gamma\varrho}\frac{\alpha}{\mu}\frac{y\,x^3 }{T\left(x^2+y^2\right)^3}
\label{eq:hea_sigma_zero_range}
\end{align}
with $\gamma^2/2 = \SI{0.02775}{\atomicunit\squared}$ the electron affinity of hydrogen, and  $\varrho=\SI{2.646}{\atomicunit}$ the effective range \citep{deJong1972} in the Standard Model.
The effective range $\varrho$, having dimensions of length, must be proportional to $a_0$. The electron affinity is defined in \citet{Bethe1949} as $\gamma^2/2 = (\mu\,\inc E)/(2 \hbar) \propto m^2 \alpha^2$, so $\gamma \propto 1/a_0$. Thus, the re-scaling in the two terms effectively cancels, and we can ignore the parametric dependence of the
$1/(1-\gamma\varrho)$ term.

We then find the cross-section for hydrogen electron attachment (reaction 3 on our Table) from the Milne relation [Eq.~(\ref{eq:Milne})] and $\mu=m$, 
\begin{align}
\sigma_{3}(x,y) &\propto \frac{ \alpha}{\mu^2}\frac{x\,y }{\left(x^2 + y^2\right)},
\end{align}
which re-scales as [Eq.~(\ref{eq:overall_scaling_factor})]
\begin{equation}
\gamma_{3,\rm DM}(T) \approx \rac{2}{-2} \, \gamma_{3,\rm SM}(\tilde{T}_a).
\end{equation}

The full Standard Model rate is given as \citep{Galli1998}
\begin{equation}
\gamma_{3,\rm SM}(T) = \SI{1.4e-18}{\epccm} \; T^{0.928} \exp\left(\frac{T}{16200}\right),
\end{equation}
and so the final, re-scaled dark rate is
\begin{align}
\gamma_{3,\rm DM}(T) =& \SI{1.4e-18}{\epccm}\, \left[\rac{0.1444}{-2.928}\right]
\nonumber\\
&\,\times T^{0.928} \exp\left(\frac{T}{16200\,\rac{2}{}}\right).
\end{align}
Comparing this result with Eq.~(\ref{eq:dark_hydrogen_rescale_recomb_rate}), we note that, while both rates use the same overall pre-factor $g(r_\alpha,r_m,r_M)$, the final re-scalings are quite different. This divergence highlights the necessity of including the $r_{\Delta E}=r_\alpha^2 r_m$ temperature re-scaling factor.

Conveniently, Eq.~(\ref{eq:overall_scaling_factor_photo}), (\ref{eq:hea_sigma_zero_range}) and $\mu=m$  are sufficient to find the re-scaling for the ${\rH}^-$ photodetachment rate, giving
\begin{equation}
\gamma_{4,{\rm DM}}
\approx
\rac{5}{} \gamma_{4,{\rm SM}}(\tilde{T}_a)\,.
\end{equation}

\subsection{Reaction 7: mutual neutralization} \label{sec:mutual neutralization}
The cross section for the mutual neutralization reaction, $\rH^- + p\rightarrow 2 \rH$, was computed from the Landau-Zener transition probability at the initial and final potential energy curve psuedo-crossing \citep{Bates1955}. Although there are two possible final states, $\rH(1s) \rH(2s \text{ or } p)$ and $\rH(1s) \rH(3s, p, \text{or } d)$, for our purposes the resulting cross section is effectively
	\begin{equation}
		\sigma_7 \propto R^2\left(\frac{k_{f}^2}{k_i^2}\right)\,F_3(\zeta),
	\end{equation}
	where $R\propto a_0$ is the internuclear distance at the potential crossing point, $k_i$ and $k_{f}$ are the momentum of the initial state and final states, with $k_i^2\propto \mu \, \KE$ and $k_f^2 \propto k_i^2+\mu \inc E = \mu \left(\KE+\inc E\right)$, $F_3(\zeta)=\zeta^{n-1}\Gamma(n-1,\zeta)-(2 \zeta)^{n-1}\Gamma(n-1,2 \zeta)$ and $\Gamma(n-1,\zeta)$ is the incomplete gamma function. The term $\zeta$ is a dimensionless ratio of momenta, which, after some simplification, can be written as $\zeta \propto k_f^{-1} (\alpha \mu \inc U)/\inc E^2$, and $\inc U$ and $\inc E$ are the zeroth order and true potential energy curve differences. In our parameter regime of interest, $\zeta \ll 1$ and a series expansion gives $F_3(\zeta)\approx\zeta$. Then we have
   
    \begin{align}
    	\sigma_7 &\propto a_0^2 \left(\frac{\KE+\inc E}{\KE}\right) \left(\frac{\alpha\,\sqrt{\mu} \, \inc U}{\inc E^2 \sqrt{\KE+\inc E}}\right) \nonumber\\
    	&\propto \alpha a_0^2 \sqrt{\mu} \frac{\sqrt{\KE + \inc E}}{\KE\, \inc E} \nonumber\\
    	&\propto  \frac{\alpha a_0^2 \sqrt{\mu}}{T^{3/2}}\frac{\sqrt{x^2+y^2}}{x^2 y^2}
    	\label{eq:mutual_neutral_scaling}
    \end{align}
	where, in the second line, we used the fact that $\inc U$ and $\inc E$ are proportional. Then, since $\inc E$ is proportional to the atomic binding energy and $\mu$ here is the proton mass, we obtain a net re-scaling of 
    \begin{equation}
		\gamma_{7,\rm DM} \approx \rac{-3}{-3} \gamma_{7,\rm SM}(\tilde{T}_a)\,.
    \end{equation}
		
\subsection{Reactions 10 and 15: Langevin and detailed balance}
\label{sec:detailed_balance_example}
An interaction without a short-range potential barrier will always occur as long as the incoming particle passes the centrifugal potential barrier and spirals into the target. These reactions can be well-treated by classical capture models, the theory of which is presented in e.g.~\citet{Hirasawa1969,Zhang2017}. In this approach, the long-range potential which governs the reaction can be written as a series in powers of the separation $R$: 
\begin{equation}
	V(R) = -\sum_n \frac{C_n}{R^n}.
\end{equation}
The leading order term in this series dominates in general, and in the Langevin (charge-induced dipole) case,  $n=4$, $C_4 = pe^2/2$ where $p$ is the isotropic polarizability of the molecule, proportional to the trace of $p_{ij}$. The reaction occurs if the collision energy exceeds the maximum of $V_{eff}(R)=V(R)+V_{\rm centrifugal}(R)$, where we  include a centrifugal term, $V_{\rm centrifugal}(R)$, to account for a non-zero impact parameter. Then, since the impact parameter is proportional to the angular momentum, we obtain the cross section in the Langevin case:
\begin{align}
\sigma_{\rm Lang}(x) 
\propto \sqrt{\frac{p e^2}{x^2 k_B\,T}} 
\propto \sqrt{\frac{a_0^3\,\alpha}{T}}\frac{1}{x},
\end{align}
where we have used the property that $p\propto a_0^3$. Then using our tools from Section \ref{sec:rate_overview_summary} and the definition $\mu \approx M$, we obtain an approximate rate re-scaling of 
\begin{equation}
\gamma_{\rm Lang,DM}(T)\approx \racx{-1}{-3/2}{-1/2} \gamma_{\rm Lang,SM}(\tilde{T}_a),
\end{equation}
which we use for both of the Langevin-type reactions: associative detachment of $\rH$ and $\rH^-$ ($\rH^-+\rH\rightarrow \rHt+e$) and $\rHt$ formation due to $\rHt^+$ ($\rHt^+ + \rH\rightarrow \rHt + p$).

The computation for the inverse reaction, $\rHt^+$ charge exchange ($\rHt + p \rightarrow \rHt^+ + \rH$), cross section uses the more general detailed balance results combined with the Langevin cross section. In this case, we have $p_{10} = \mu_{10} v_{10} \approx \mu_{15} v_{15} = p_{15}$ (where the labels $10$ and $15$ correspond to those reactions), since $\mu_{10}=(m_{\rHt^+} \,m_{\rH})/(m_{\rHt^+} +m_{\rH})\approx (2 \,M^2)/(3\,M) \approx (m_{\rHt} M)/(m_{\rHt}+M) \approx \mu_{15}$. With the standard assumption that the internal degrees of freedom $g_{10}$ and $g_{15}$ stay constant, we have $\sigma_{15}\propto\sigma_{10}$, and we only need the temperature re-scaling, which, from reaction 10, is proportional to the hydrogen binding energy $\eh$.

\subsection{$\rHt$ Dissociation Scaling} 
\label{sec:h2_diss}
	We approximate the $\rHt$ dissociation reaction, $\rHt + \rH \rightarrow 3\rH$, at the lowest level as a hard-sphere collision. Then, since the hard-sphere impact parameter, $b$, scales as $a_0 \propto (\alpha m)^{-1}$, we have 
	\begin{equation}
		\sigma_{\rm diss} \propto b^2 \propto \left(\frac{1}{\alpha m}\right)^2.
	\end{equation}
	The dominant binding energy is simply the dissociation energy, $D_e$ which depends on $\alpha^2 m$, so $r_{D_e}=\rac{2}{}$. The overall re-scaled rate is then given by
	\begin{align}
		\gamma_{\rm diss,DM}(T) &\approx  \left(\sqrt{\frac{r_{D_e}}{\rx{}}}\right) \left(\rac{-2}{-2}\right)\; \gamma_{\rm diss}(T/r_{D_e})\\
		&=\racx{-1}{-3/2}{-1/2}\; \gamma_{\rm diss,SM}(\tilde{T}_a),
		\label{eq:h2_diss_scaling}
	\end{align}
	where the first term in brackets comes from the velocity average (with the approximation $\mu=m_{\rHt}=M$) and the second from the cross section. 
	
	Note that there is a major limitation to this re-scaled rate: the hard-sphere approximation introduces significant simplifications into the cross section parametric dependence that are not necessarily justified. As \cite{Martin1996} points out, the $\rHt+\rH \rightarrow 3\rH$ reaction rate requires accounting for all of the possible energy level transitions and populations, including not just collisions but also spontaneous dissociation via tunneling from quasi-bound states. This introduces myriad locations for the various parameters to creep in, including, at a minimum, the resulting temperature dependence in the cross section. 

\subsection{Three-Body-Reactions}
\label{sec:Three-Body}

We re-scale the reaction rate for the 
following three-body reactions, taken from \cite{Yoshida_2006}:
\begin{align}
3 \rH &\rightarrow \rHt + \rH\\
\rHt + 2 \rH &\rightarrow 2\rHt \\
2 \rH + \rH^+ &\rightarrow \rHt + \rH^+\\
2 \rH + \rH^+ &\rightarrow \rHt^+ + \rH.
\end{align}
The re-scaling is common to all four reactions, and can be derived from the principle of detailed balance (see, e.g.~\cite{Flower2007}). 
The chemical equilibrium relation gives
\begin{equation}
\gamma_a n_{\rH}^2 = \gamma_d n_{\rHt},
\end{equation}
where $\gamma_a$ is the rate coefficient for three-body association reaction
and $\gamma_d$ is the rate coefficient for the inverse process. Then, using the Saha equation:
\begin{equation}
n(\rHt) = n^2(\rH) \frac{Z_{\rHt}}{Z_{\rH}^2} \left(\frac{1}{\pi M T}\right)^{3/2} e^{D_{e,D}/T},
\end{equation}
where $Z= \sum_i g_i e^{-E_i/T}$ is the partition function with $g_i$ the statistical weight. 

Substituting, and noting that the partition function for atomic hydrogen (assumed predominantly in the ground state) is a constant (whose value is given as $1$ in \cite{Flower2007} and $2$ in \cite{Forrey2013}):
\begin{equation}
\gamma_{a,\rD}(T) = \gamma_{d, \rD}(T) \frac{2}{Z_{\rHt}} \left(\frac{1}{\pi M T}\right)^{3/2} e^{D_{\rm e}/T}.
\end{equation}
\cite{Flower2007} gave $Z_{\rHt} \approx .028 T$, and at the low temperatures where these reactions are significant, this scales as the rotational energy, $T \rightarrow \tilde{T}_r$. We substitute $z_{\rHt}(\tilde{T}_r) = \rcx{}{-1} z_{H_2}(T_a)$ and apply the scaling of $\gamma_{\rm diss,DM}$ from Section \ref{sec:h2_diss} to substitute $\gamma_{\rm diss,DM}(T) = \racx{-1}{-3/2}{-1/2} \gamma_{\rm diss,SM}(\tilde{T}_a)$, yielding
\begin{align}
\gamma_{a,\rD}(T) &=\racx{-4}{-4}{-1}\nonumber\\ &\times\frac{2}{Z_{\rHt}(\tilde{T}_a)}\left(\frac{1}{\pi m_{\rH} \tilde{T}_a}\right)^{3/2} e^{D_e/\tilde{T}_a} \gamma_{\rm diss,SM}(\tilde{T}_a)\\
 &= \racx{-4}{-4}{-1}\gamma_{a,\rm SM}(\tilde{T}_a).
\end{align}

The reaction rates are taken from \cite{Yoshida_2006} and \cite{Krstic2003}. This derivation can be applied to the final reaction of the four (which involves a charge exchange) by considering the Saha factors between $\rHt^+$ and $\rHt$ as well as between $\rH$ and $\rH^+$. 
		
\section{Validity of results} \label{sec:validity}
There are several limitations to the results presented above, some of which are based on assumptions we have made, whereas others follow from the dependence on Standard Model chemistry. Therefore, many of the results so far presented are only valid in certain regions of the parameter space and we discuss the reasons and regions here.

\subsection{Limitations inherited from Standard Model results} \label{sec:valid_reactions}
The largest source of inaccuracy of our results follows from our dependence on the Standard Model rates. Many of the reactions, especially those considered ``unimportant" (in Table \ref{tab:reactions_table_part3}) are poorly known, while others are only known for specific cases. For example, the reaction $\rH^- + p \rightarrow 2\rH $ was only known to within an order of magnitude until recently \citep{Glover2015}, while higher vibrational levels in the $\rHt + e\rightarrow \rH + \rH^-$ reaction were not thought to have a significant contribution to the net rate until \citet{Capitelli2007}. 
	
Most of the reactions listed in Table \ref{tab:reactions_table_part1} are obtained from older cross sections that may not account for these corrections and special cases found more recently. However, we expect that as long as the approximations used in the older rates remain reflective of the underlying physics, the rate re-scalings obtained will remain valid, as demonstrated in Figures \ref{fig:ldl_scaled_cooling} and \ref{fig:recomb_comparison}. 
	
Additionally, the majority of reaction cross sections listed in section \ref{sec:reactionrates} are analytic fits to numerical computations or even experimental data. As such, they are only defined for certain temperature regimes. For example, the reaction $\rH+p\rightarrow \rHt^++\gamma$ has a reaction rate defined on the interval $\SI{1}{\kelvin}\le T \le \SI{32000}{\kelvin}$, but is undefined elsewhere. Extrapolation past those temperature limits can result in significant divergences, especially with some of the more accurate fits that have strong exponential terms, like the rovibrational cooling rates mentioned in Section \ref{sec:cooling}. These temperature limits will scale with the dark parameters but extra care should still be taken when exploring more extreme parameter space ranges. 
	
A more subtle issue exists for radiative reactions because the gas and photons need not share a temperature. In principle, the reaction rates should then depend on both the gas and photon temperature, but all of our rates depend on only one of these temperatures. \citet{CyrRacine2013} points out that in particular the standard calculation of the recombination coefficient neglects stimulated recombination, which explicitly depends on the photon temperature. That work found a change of at most a factor of a few in the recombination coefficient due to this effect when $T_g = 0.01 T_{\gamma}$. In the particular case of hydrogen recombination, this issue is relevant only for weakly coupled dark matter models, since otherwise the gas and radiation are thermally coupled at recombination. For the problem of halo formation \citep{KROME2014}, the matter and radiation do not in general share a temperature, but even so standard calculations consider each rate to be a function of only one temperature. 
	
A final limitation derives from several of the rates listed here depending on the assumption from the Standard Model that $m\ll M$. Trivially, in many cases we approximate the reduced mass $\mu=(m\,M)/(m+M)\approx m$ for the $e$, $p$ system, or $\mu\approx m_{\rH} \approx M$ for the $H$, $H$ (or $p$) system, which can be easily adjusted as $m\rightarrow M$.  On the other hand, in section \ref{sec:basic_tools} we assumed stationary nuclei, with electronic motion resulting in minor corrections at most. This no longer holds as $m\rightarrow M$, as we can no longer use the Born-Oppenheimer approximation. This affects both the fundamental wave functions and cross section calculations. For example, the derivation for the hydrogen electron attachment reaction ($\rH+e\rightarrow \rH^- + \gamma$) assumed a massive stationary, screened proton captures a free electron. In the extreme case, $m\approx M$, the physics should become much closer to positronium production, $e^- + e^+\rightarrow Ps$ or radiative association of $\rH$ and $p$, $\rH+p\rightarrow \rHt^+ + \gamma$, with the corresponding re-scaling, but where and how the transition would occur is beyond the scope of this work. Even the basic hydrogen atomic physics are modified; with a significantly larger electron mass, the energy eigenvalues would have a significant dependence on the nuclear (proton) mass \citep{Poszwa2016}.

\subsection{Comparison with deuterated chemistry}
\begin{figure}[htbp!]
\includegraphics[width=0.49\textwidth]{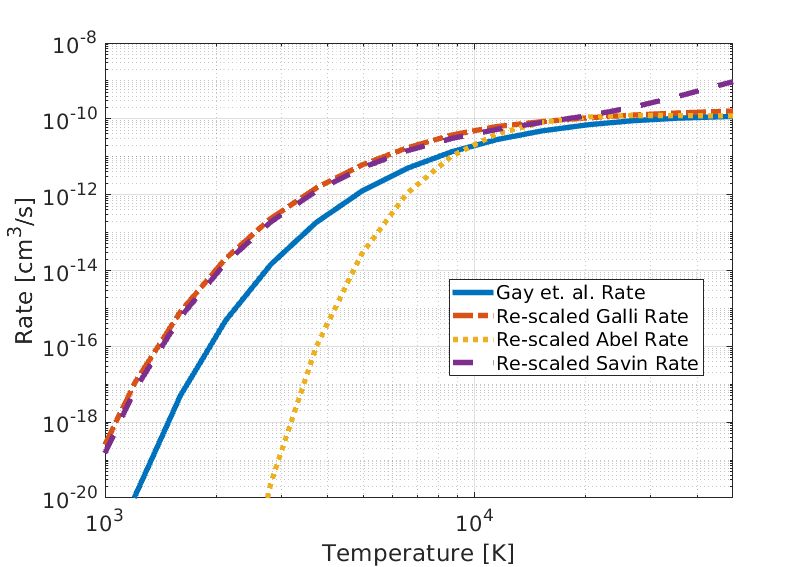}
\caption{$\rD_2+\rD^+\rightarrow \rD_2^+ + \rD$ reaction rate and re-scaled $\rH$ rates for comparison. Re-scaled $\rHt + \rH^+\rightarrow \rHt^+ + \rH$ reaction rates are from \citet{Galli1998,Abel1997,Savin2004} and the deuterium rate is from \citet{Gay2011}. The deuterium rate is clearly comparable to the re-scaled hydrogen rates to the extent that they are comparable to each other. }
		\label{fig:deut_comparison}
\end{figure}

As mentioned in the introduction, we can compare our rates with those obtained for deuterium reactions. \citet{Gay2011} has compiled a significant list of deuterium reaction rates; while the majority are computed using the mass re-scaling described in Appendix \ref{sec:mass_scaling}, or assumed identical to the hydrogen rate (e.g. $\rD^+ +e \rightleftharpoons \rD + \gamma$), one is based directly on experimental data. The reaction $\rD_2+\rD^+\rightarrow \rD_2^+ + \rD$ has a rate derived from the cross section listed in \citet{Wang2002} and given by 
\begin{align}
&\gamma_{D_2+D^+\rightarrow D_2^+ + D} 
\nonumber\\
=&\, \SI{2.46e-9}{\centi\meter\cubed\per\second}\, T_3^{-0.439} \exp\left(\frac{-{30500}}{T}\right).
	\label{eq:gay_deut_rate}
\end{align}
While this appears significantly different to our re-scaled result from \citet{Galli1998}, $\gamma_{15,\rm DM} \approx \SI{2.12e-10}{\centi\meter\cubed\per\second} \,\exp\left(-21050/T\right)$, the discrepancy is not fatal to our argument. \citet{Savin2004} points out that there is significant variation between the reported rates for the $\rHt + \rH^+\rightarrow \rHt^+ + \rH$ reaction, with sometimes drastically different values and temperature behavior even between rates based on the same source data. While the rates may differ significantly, they share common features, including an exponential drop off dependent on the threshold energy at low temperatures, and minimal temperature dependence in the \SIrange{e4}{e4.5}{\kelvin} range.  In Figure \ref{fig:deut_comparison}, we plot several re-scaled hydrogen rates (from \citet{Galli1998,Abel1997,Savin2004}), as well as the rate from Gay, to demonstrate that the rates are sufficiently similar, especially at higher temperatures. 
	
As stated, the remaining pure deuterium reactions in \citet{Gay2011} not identical to hydrogen rates have been computed using the mass re-scaling method described in section \ref{sec:mass_scaling}, with a minor change: in Eq.~(\ref{eq:simple_mass_scaling}), $a_{2,\rm known}\rightarrow a_{2,\rm known}+0.5$. To the extent that those rates have been used in many astrochemistry calculations, this further validates our approach of re-scaling known Standard Model rates, and comparison of Eq.~(\ref{eq:simple_sigma}) with Eq.~(\ref{eq:overall_rescaled_rate}) and Eq.~(\ref{eq:overall_scaling_factor}) reveals that the simplified version is equivalent to $g(\ra{},\rc{},\rx{})=r_{\mu}^{-1/2-a/2}$ and $\inc E$ at least approximately independent of $\mu$. Since $\mu\propto M$ for these cases, this holds reasonably well for those reactions with simple cross sections and atomic binding energy, for example, mutual neutralization, $\rH^{-}+p\rightarrow 2 \rH$. Our more complete re-scaling procedure ($\gamma_{7,\rm DM}\propto \rx{-0.5}\gamma_{7,\rm SM}$) gives equivalent results to the mass re-scaling ($\gamma_{7,\rm DM}\propto \rx{-0.487}\gamma_{7,\rm SM}$). 
	
\begin{figure}[htbp!]
		\includegraphics[width=0.49\textwidth]{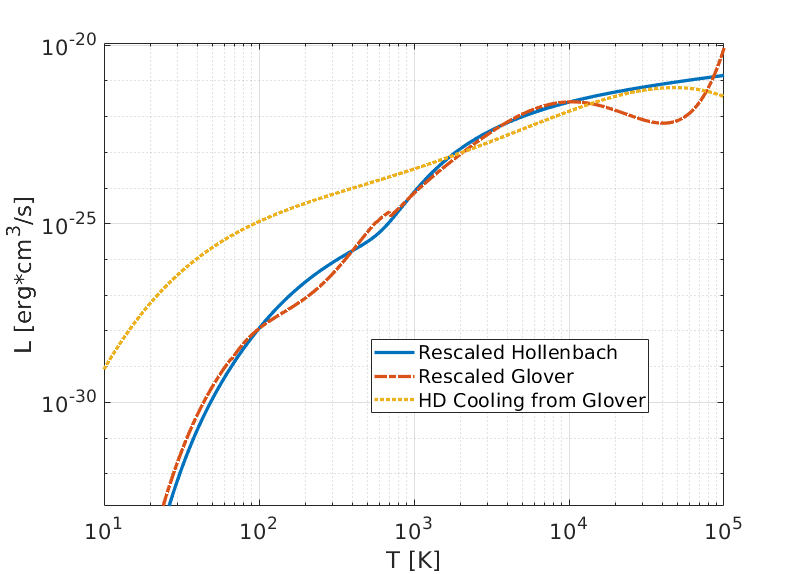}
		\caption{Demonstration of the inability to utilize the $\rH\rD$ molecule in validation. While the re-scaled \citet{Hollenbach1979} and \citet{Glover2008} cooling curves are similar to the $\rH\rD$ cooling curve in the high temperature region, at low temperatures the allowed $J\pm1$ transitions enable $\rH\rD$ to have an orders of magnitude higher cooling rate.} 
		\label{fig:HD_Comparison}
	\end{figure}

Lastly, we consider the cooling rates and demonstrate why we do not validate using $\rH\rD$ chemistry. Unfortunately, we were unable to find a $\rD_2$ cooling function based on experimental data or theoretical calculations and can only compare our re-scaled rates to $\rH\rD$ cooling rates, setting $M=1.5 m_p$. However, $\rH\rD$ has significantly different behavior, as compared to $\rHt$, due to its mass asymmetry-induced dipole moment and allowance of $J\pm1$ transitions. Thus, the cooling comparison breaks down in the low temperature regime where those aspects become relevant. We demonstrate this issue in Figure \ref{fig:HD_Comparison}, where we plot the re-scaled \citet{Hollenbach1979} and \citet{Glover2008} $\rHt$ cooling rates versus the HD cooling rate from \citet{Glover2008}. Note that while the differences between the $\rH\rD$ curve and the re-scaled curves are small in the high temperature region, at low temperatures the $\rH\rD$ curve is orders of magnitude higher. 
	
\subsection{Additional reactions} \label{sec:additional_reactions}

The minimal reaction list of \cite{Galli1998} informs our reaction list. However, we neglect several of these reactions (briefly listed in Table \ref{tab:reactions_table_part3}). This list includes three-body reactions and reactions deemed irrelevant by \cite{Galli1998}. Since different reactions are re-scaled by different amounts, one might worry that reactions which are insignificant in the Standard Model could contribute. There are three distinct reasons a reaction might fail to contribute:
\begin{enumerate}
    \item It may be out-competed by a different channel involving identical reactants, for example $\gamma_6$ compared to $\gamma_7$.
    \item It may be suppressed below some threshold energy which is large compared to competing reactions. For example reaction 18 requires a photon with enough energy to excite the electronic degrees of freedom of the $\rm H_2$ molecule, but except in the case of a cold gas exposed to an external Lyman-Werner flux (which we never encounter) collisional dissociation mechanisms will efficiently destroy the molecules at a much lower temperature. 
    \item It may be suppressed by the relative abundances of the species involved. For example, reaction 10 dominates over reaction 13 because in general there is more neutral hydrogen than $\rH_2$.
\end{enumerate}
For the first case, since the basic energy and angular momentum structure of the molecule is preserved under re-scaling, whatever physical principle (i.e.~selection rules, charge-dipole interactions) prefers one channel over another continues to operate. In the second case, the sharp cutoff in the reaction rate below the threshold energy cannot be overcome by parametric re-scaling of the rate coefficient. In the third case, serious care is required because changes to the abundances or the reaction rates can alter the relative importance of the various reactions. Note also that these categories are not completely disjoint. In the example mentioned for (2), if one encounters a case where $\rH_2$ molecules and energetic photons coexist, then the considerations of case (3) might still prevent the reaction from contributing.

We have re-scaled and included all reactions omitted due to (3) except for reaction 11, $\rHt^+ + e \rightarrow 2\rH$. This reaction is outcompeted by reaction 10, $\rHt^+ +\rH \rightarrow \rHt + p$,  in all cases considered because we always find $x_\rH \gg x_e$ by the time $\rHt^+$ forms.

	\begin{table}
		\scriptsize
		\centering
			\begin{tabular}{l l l }
				\toprule
				\# & Reaction & Justification\\
				\midrule
				6 & $\rH^- + p \rightarrow \rHt^+ + e$&1 \\
				11 & $\rHt^+ + e \rightarrow 2 \rH $ & 3   \\
				12 & $\rHt^+ + \gamma \rightarrow 2p + e$ &1\\
				14 & $\rHt^+ + \rH\rightarrow \rH_{3}^+ + \gamma$ &1\\
				16 & $\rHt + e \rightarrow \rH + \rH^-$ &2 \\ 
				17 & $\rHt + e \rightarrow 2 \rH + e$& 1,2\\
				18 & $\rH_{2} + \gamma\rightarrow \rHt^+ + e$&2 \\
				19 & $\rH_{3}^+ + \rH\rightarrow \rHt^+ + \rHt$  &2\\
				  21 & $\rHt + \rH^+\rightarrow \rH_{3}^+ + \gamma$& 1\\
				22 & $\rH_{3}^+ + \gamma\rightarrow \rHt^+ + \rH$& 2 \\
				\bottomrule
			\end{tabular}
			\caption{Reactions from Galli and Palla that are not part of the minimal model and were not included in the re-scaling procedure with the justification for omission, as enumerated in the text. 
			\label{tab:reactions_table_part3}}
			\begin{tablenotes}
				
			\end{tablenotes}
	\end{table}

\section{Conclusion} \label{sec:conclusion}
In order to extend the study of atomic dark matter models and determine their impact on the formation and evolution of large-scale structure, galactic halos, and compact object, we have extended previous studies \citep{Rosenberg2017,Buckley2018,Boddy2016} by calculating the chemistry of dark molecular hydrogen. As dark molecular hydrogen processes dominate the chemistry at low temperature, this work provides an essential tool for studying the formation and evolution of low-mass halos with much lower virial temperatures, and the end point of cooling processes within halos. 

We have provided some basic tools for working with dark molecules, including the re-scaling for several key atomic and molecular quantities (Table \ref{tab:base_quantity_scaling}), and the procedure for obtaining approximately re-scaled rates when the physics of the analogous Standard Model interaction is known (e.g. Eq.~(\ref{eq:basic_gamma_rot_scale}), Eq.~(\ref{eq:overall_rescaled_rate}), and  Eq.~(\ref{eq:overall_scaling_factor})). We have applied these tools to two important areas in molecular chemistry, thermal processes and chemical reactions, and determined how the rates in a minimal subset of these processes can be re-scaled to obtain expressions valid for a wide range of dark matter parameters (Eq.~(\ref{eq:lte_rot_scale}), Eq.~(\ref{eq:lte_vib_scale}), Eq.~(\ref{eq:ga_full_sum}) and Table \ref{tab:reactions_table_part1}). Importantly, these re-scaled rates are only as accurate as the original Standard Model rates and neither correct nor transcend any limitations or assumptions therein, as demonstrated with our comparison with deuterated reactions (Eq.~(\ref{eq:gay_deut_rate}) and Figures \ref{fig:deut_comparison} \& \ref{fig:HD_Comparison}). 
	
Additionally, we note that many astrochemical databases and chemical models either neglect deuterium reactions entirely, or when present either use the hydrogen rate directly or use the simplified mass re-scaling method from Appendix \ref{sec:mass_scaling} (e.g. the \code{deuspin} chemical network in the KIDA database, from \citet{Majumdar2016}). Therefore, at a minimum we expect the re-scaling method presented here to provide improved rates for deuterium reactions that can easily be included in said models.
	
We have provided the first calculations of dark molecular processes with sufficient accuracy to be used in simulations. These are used in our companion papers to derive cosmic abundances of various states (ionized, atomic, and molecular hydrogen) \citep{Gurian2021} and to model the low-virial temperature halo formation in analogous fashion to the minihalo formation in Standard Model \citep{Ryan2021}. Eventually, these rates and the re-scaling procedure can be used to validate phenomenological models used in large-scale structure simulations for studying the formation and evolution of dark-matter halos, which increasingly depend on full numerical calculations that depend on chemistry rates that we study here. 

At the core of the atomic-dark-matter halo, the dark-matter density may reach high enough values so that the re-scaling of other process rates, especially many-body processes such as $\rH_3$ reactions, might also be necessary. Of course, to get the dark-chemistry rates, one can also take the parallel approach of computing the interaction rates from the first-principle quantum calculations. We leave these computations for future work.
		
\acknowledgments
Funding for this work was provided by the Charles E. Kaufman Foundation of the Pittsburgh Foundation. We thank the anonymous referee for providing a truly excellent and professional report, which contributed significantly to the depth of analysis presented in the revised version.

\appendix

\section{Brief Review of Mass Re-scaling}
\label{sec:mass_scaling}
Applications of chemistry in astrophysics frequently require chemical reaction rates that are difficult to compute from first-principles, and are required in temperature and density regimes inaccessible in laboratory experiments on Earth. Sometimes, rates may be known for other isotopes or related species and in these cases a common approach in the literature (see e.g. \citet{Stancil1998,Walker2014}) is similar to what we present in this work: re-scaling the known rate by a ratio of the masses.

Assuming the known (endoergic) reaction rate can be approximated by the form
\begin{equation}
	\gamma_{\rm known} = a_1 \; T^{a_2}\;\exp\left(\frac{a_3}{T}\right),
	\label{eq:endoergic_rate}
\end{equation}
with temperature $T$ and constants $a_i$, then in the mass-scaled rate the coefficient $a_1$ takes the form
\begin{equation}
	a_{1,\rm new} = a_{1,\rm known} \left(\frac{\mu_{\rm known}}{\mu_{\rm new}}\right)^{a_{2,\rm known}},
	\label{eq:simple_mass_scaling}
\end{equation}
where $\mu_{\rm known}$ ($\mu_{\rm new}$) is the reduced mass of the known (new) reaction. All other terms remain the same and the same formula holds for exoergic rates. For example, the rate coefficient for deuterium mutual neutralization, $\rD^- + \rD^+\rightarrow 2\rD$ can be approximately obtained by re-scaling the rate for the regular hydrogen reaction, $\rH^- + p \rightarrow 2 \rH$. The hydrogen reaction is known to be of the form of Eq.(\ref{eq:endoergic_rate}) with $a_1=\SI{1.4e-7}{\centi\meter\cubed\per\second}$ and $a_2=\num{-0.487}$ \citep{Stancil1998}. Using $\mu_\rD\approx\SI{1}{\giga\electronvolt}$ and $\mu_\rH\approx\SI{0.5}{\giga\electronvolt}$, the mass re-scaling procedure suggests the corresponding deuterium process is proportional to $a_{1,\rD}=a_{1,\rH}\left(\mu_\rH/\mu_\rD\right)^{a_{2,\rH}}\approx \num{1.4}\,a_{1,\rH} = \SI{1.96}{\centi\meter\cubed\per\second}$. 

The derivation of this re-scaling relationship is straightforward \citep{Walker2014}. Assume the reaction cross section depends on the collision velocity $v$ as
\begin{equation}
	\sigma(v) = B v^a,
	\label{eq:simple_sigma}
\end{equation}
where $B$ is a mass independent constant with units $[{\rm length}]^{2-a}[{\rm time}]^a$. The rate is given by the thermal average,
\begin{equation}
	\gamma=\langle\sigma\,v\rangle = \left(\frac{2}{\pi}\right)^{1/2}\left(\frac{\mu}{k_B T}\right)^{3/2} \int_0^{\infty} \sigma(v) \exp(-\mu v^2 /2 k_B T) v^3 dv\,.
\end{equation}
The integral can be performed exactly, giving
\begin{equation}
	\gamma = \frac{B}{\sqrt{\pi}} 2^{\frac{a+3}{2}}\, \left(\frac{k_B T}{\mu }\right)^{\frac{a+1}{2}} \Gamma \left(\frac{a}{2}+2\right)= A(a) \left(\frac{T}{\mu}\right)^{b}.
\end{equation}
Then re-scaling $\mu$ gives back the expression found in Eq.~(\ref{eq:simple_mass_scaling}), where $a_1 = A(a) \mu^{-b}$ and $a_2=b=(a+1)/2$. Note that the exponential cutoff in Eq.~(\ref{eq:endoergic_rate}) not derived in \citet{Walker2014} can arise naturally within this formalism in at least two ways. First, by imposing a lower energy cutoff in the integration: 
\begin{align}
    \gamma &=  \left(\frac{2}{\pi}\right)^{1/2}\left(\frac{\mu}{k_B T}\right)^{3/2} \int_{v_{min}}^{\infty} \sigma(v) \exp(-\mu v^2 /2 k_B T) v^3 dv \\
    &=\frac{B}{\sqrt \pi}2^{\frac{a + 3}{2}}\left(\frac{k_B T}{\mu}\right)^{\frac{1+a}{2}}\Gamma\left(2 + \frac{a}{2}, \frac{\mu v^2_{min}}{2 k_b T}\right),
\end{align}
and approximating the incomplete gamma function in $1/T$ by an exponent in $-1/T$ (which is exact for $a = -2$). Second, if there is no minimum energy cutoff in the forward reaction (i.e.~Langevin reactions), an exponential term will appear in the reverse reaction due to the detailed balance factor. This can give us some confidence that the form Eq.~(\ref{eq:simple_mass_scaling}) is physically well motivated and could be reasonable to re-scale empirical fits of this form over a wide temperature range.

Unfortunately, Eq.~(\ref{eq:simple_mass_scaling}) is insufficient for computing dark reaction rates for two reasons. First, it clearly cannot capture the effect of allowing $\alpha$ and $m$ to vary from their Standard Model values. Second, even in the case where $\ra{}=\rc{}=1$, the $B$ term in Eq.~(\ref{eq:simple_sigma}) is dimensionful, and often contains factors of $\mu$ that are not accounted for. Likewise, the dimensionful $a_2$ and $a_3$ terms in Eq.~(\ref{eq:endoergic_rate}) can depend on $m$, $M$, and $\alpha$, significantly changing the rate's temperature dependence. The re-scaling procedure we have presented in this paper overcomes these limitations.

\section{Quadrupolar spontaneous emission in Atomic Units}
\label{app:AU}
The quadrupolar spontaneous emission rate is usually given in atomic units in the literature. In these units, $e^2=4\pi\epsilon_0 = \hbar \equiv 1$, masses are measured in $m_e\equiv 1$, and energy is measured in units of the Hartree energy, $\eh=\alpha^2 m_e c^2=1$. With these definitions, $c=1/\alpha$, and the unit of time is $\hbar/\eh$.

When the energies and quadrupole moments are given in atomic units, the rate for a standard model hydrogen atom to transition from a state with (vibrational, rotational) quantum numbers ($\nu^{\prime},J+2$) to the state with ($\nu,J$) is \citep{Turner1977,Flower2007}
\begin{align}
\label{eq:quadAU}
A(\nu, J\leftarrow \nu^{\prime},J+2)=&\left(\SI{1.43e4}{\per\second}\right)(E_{\nu^{\prime},J+2}-E_{\nu,J})^5\frac{3(J+2)(J+1)}{2(2J+5)(2J+3)}Q^2_{\nu^{\prime},J+2;\nu,J}\nonumber\\
=&\frac{1}{60c^5}(E_{\nu^{\prime},J+2}-E_{\nu,J})^5\frac{3(J+2)(J+1)}{2(2J+5)(2J+3)}Q^2_{\nu^{\prime},J+2;\nu,J}
\end{align}
The rates for other transitions, with $\inc J=0, -2$, have the same form. The numerical pre-factor in the top line can be obtained from the second line using the conversion to S.I. units from the unit of time in atomic units, $\hbar/\eh$,  ($\hbar/\eh=\SI{2.418884e-17}{\second}$ in S.I. units), $c=137$, and the factor of 60 (note the typo in \citet{Flower2007}). 

From Eq.(\ref{eq:quadAU}), the re-scaling relationships needed to approximate dark matter transition rates can be determined as a change in units from Standard Model atomic units to dark matter atomic units. Given a numerical value of some transition rate in the Standard Model, the parametric dependence comes from changing units in the factor of $\eh/c^5=\eh\alpha^5$ in the pre-factor in Eq.~(\ref{eq:quadAU}), and from the fundamental constants appearing in the energy difference of vibrational or rotational modes when those energies are expressed in atomic units. Since the quadrupole moment is proportional to $a_0^2$, there is no additional parametric dependence from $Q^2$. That is, suppose a Standard Model quadrupole in S.I. units is $Q\equiv\tilde{Q}a_0^2$, where $\tilde{Q}$ is value of the quadrupole in Standard Model atomic units. Then the dark matter quadrupole can be obtained by $Q^{\rm DM}=Q\left(a_{0,{\rm DM}}/a_0\right)^2$. Using dark atomic units on the left hand side, $\tilde{Q}^{\rm DM}=Q\left(1/a_0\right)^2$, or $\tilde{Q}^{\rm DM}=\tilde{Q}$.

Finally, then, the dark matter Einstein $A$ coefficient for transitions between rotational levels can be estimated by the re-scaling
\begin{align}
A_{\rm quad, rot\, DM.}=&\left[\left( \frac{\alpha_{\rm DM}}{\alpha_{\rm SM}}\right)^5\frac{E_{\rm h,\,DM}}{E_{\rm h\,SM}}\right]\left[\left(\frac{\inc E_{\rm rot,DM}}{\alpha_{\rm DM}^2mc^2}\right)\left(\frac{\alpha_{\rm SM}^2m_ec^2}{\inc E_{\rm rot,SM}}\right)\right]^5\; A_{\rm quad, rot\;SM}\\\nonumber
=&\left[\frac{r_{\alpha}^7r_m^6}{r_M^5}\right]\; A_{\rm quad,rot,\;SM}
\end{align}
and, for vibrational transitions,
\begin{align}
A_{\rm quad, vib\, DM.}=&\left[\left( \frac{\alpha_{\rm DM}}{\alpha_{\rm SM}}\right)^5\frac{E_{\rm h,\,DM}}{E_{\rm h\,SM}}\right]\left[\frac{\inc E_{\rm vib,DM}}{\alpha_{\rm DM}^2mc^2}\frac{\alpha_{\rm SM}^2m_ec^2}{\inc E_{\rm vib,SM}}\right]^5\; A_{\rm quad, vib\;SM}\nonumber\\
=&\left[\frac{r_{\alpha}^7r_m^{7/2}}{r_M^{5/2}}\right]\; A_{\rm quad,vib\;SM}\,.
\end{align}
These expressions agree with those obtained in the main text.

\section{The Dark Ortho- to Para-Hydrogen Ratio}\label{app:orthopara}
In low-temperature, low-density hydrogen gas, the ratio of ortho- (nuclear spin $I=1$, corresponding to rotational level $J=1$) to para- ($I,J=0$) forms of molecular hydrogen plays a key role in rovibrational cooling, due to the differing energies of the lowest transition \citep{Glover2008}. In the Standard model, this ratio is set by the collisional processes $\rHt+p$ and $\rHt+\rH$, with the proton collision dominating in primordial halos, and is generally assumed to be 3 to 1 \citep{Glover2008,Lique2012}. Radiative transitions do not contribute, as their rate of $\Gamma\approx\SI{e-21}{\per\second}$\citep{Pachucki2008} is several orders of magnitude less than the equivalent proton rate, $n_p k_{p,O\rightarrow P}(\SI{50}{\kelvin})\approx\SI{e-10}{\per\second}$, assuming $n_p=\SI{1}{\per\centi\meter\cubed}$\citep{Gerlich1990}. In the dark sector, however, all three interaction rates depend on the dark parameters, and their relative dominance need not match the Standard-Model case. In this section, we study the re-scaling of these rates relevant for setting the ortho-para ratio.

\subsection{Collisional and radiative scaling} \label{app:op_collrad}
The collisional reactions are straightforward to re-scale by using the results in  \citet{Gerlich1990}. 
First, we approximate the $\rHt+p$ cross section using the Langevin model and reuse our results from Section \ref{sec:detailed_balance_example}. The overall rate re-scales as $\racx{-1}{-3/2}{-1/2}$. The energy scale associated with the transition is no longer $\eh$, however, and is instead $\erot$, giving the total re-scaling of 
\begin{equation}
    \gamma_{{\rm DM},j \rightarrow j'}(T) = \racx{-1}{-3/2}{-1/2} \gamma_{{\rm SM}, j\rightarrow j'}(\tilde{T}_r) \label{eq:op_p_rate_scale}.
\end{equation}
The $\rHt-\rH$ collision rate for a rotational transition has already been re-scaled in Section Eq.~\ref{eq:basic_gamma_rot_scale},
\begin{equation}
    \gamma_{{\rm DM},j \rightarrow j'}(T) = \racx{-1}{-1}{-1} \gamma_{{\rm SM}, j\rightarrow j'}(\tilde{T}_r) \label{eq:op_H_rate_scale}.
\end{equation}

The radiative transition is more complicated, involving several corrections and effects beyond the spin-orbit interaction, but \citet{Pachucki2008} provides an excellent analytic guide. The process is composed of two transitions, the $J+I=2 \rightarrow 0$ transition with rate
\begin{equation}
    \Gamma_2 \approx \frac{1}{120}\alpha \erot^5 \left(\frac{g_p R_0}{M}\right)^2
    \label{eq:op_gamma2}
\end{equation}
and the $J+I=1 \rightarrow 0$ transition, with rate
\begin{equation}
    \Gamma_1 = \frac{2}{9}\alpha \erot |\langle \Sigma | Q^k | \Pi^k\rangle |^2.
\end{equation}
Here, $g_p$ is the proton g-factor (independent of the dark parameters), $R_0$ is the average inter-proton distance (scales as $a_0$), and the matrix element $\langle \Sigma | Q^k | \Pi^k\rangle$ can be written in terms of dimensionless functions $F_i=F_i(m \alpha R_0)$ as 
\begin{align}
    \langle \Sigma | Q^k | \Pi^k\rangle &= \frac{\erot^2 R_0}{2 M}\left[\left(\frac{g_p}{2}-1\right) - \frac{9}{2}\frac{g_p F_1}{(m \alpha R_0)^3}\right] - \frac{\erot}{M^2 R_0}\left[\left(g_p-1\right)- g_p F_3 + \frac{9}{4}\frac{g_p F_2}{(m \alpha R_0)^3}\right].
\end{align}

Using the definitions from Section \ref{sec:basic_tools}, and noting that the re-scaling of the quantity $m \alpha R_0$ is $1$, a little algebra clearly shows that $\Gamma_2$ and $\Gamma_1$ both re-scale the same way, with the angular-momentum-averaged rate, $\Gamma_{\rm avg} = (5 \Gamma_2 + 3 \Gamma_1)/9$ then simply re-scaling as 
\begin{equation}
    \Gamma_{\rm avg, DM} = \racx{9}{8}{-7} \Gamma_{\rm avg, SM}. \label{eq:op_dark_gamma}
\end{equation}

\subsection{The ortho-para ratio} \label{app:op_ratio}
As shown in Appendix~\ref{app:op_collrad}, the $\rHt+\rH$ and $\rHt+p$ ortho-para conversion rates [Eqs.~(\ref{eq:op_p_rate_scale})-(\ref{eq:op_H_rate_scale})] have a substantially weaker re-scaling than the radiative conversion rate [Eq.~(\ref{eq:op_dark_gamma})]. As such, all three need to be considered when calculating the dark ortho-para ratio.

Assuming radiative transitions remain irrelevant, the behavior of the ortho-para ratio will depend on the free proton fraction, as is true in the Standard Model. Effectively, well below the $\rHt-\rH$ activation temperature ($\tilde{T}_r=\racx{-2}{-2}{}T\sim\SI{5000}{\kelvin}$ \citep{Lique2012}), the ortho-para ratio can be approximated using the expression from \citet{Gerlich1990} at re-scaled temperature $\tilde{T}_r$,
\begin{equation}
    K_{\rm DM}(T) \approx \begin{cases}
        \num{9.35}\exp(\frac{-169.4}{\tilde{T}_r}) & \tilde{T}_r<\SI{180}{\kelvin} \\
        3 & \tilde{T}_r\ge \SI{180}{\kelvin}.
    \end{cases}
\end{equation}
Near the activation temperature, and if $n_{p}/n_{\rH}\gtrsim \num{e-3}\sqrt{\rc{}/\rx{}}$, the $\rHt+p$ reaction dominates to keep the ortho-para ratio 3 to 1. If not, the ratio approaches the $\rHt-\rH$ value, approximately \num{7}, as the temperature increases. Note, however, that the discrepancy between the fully-ortho cooling rate and the fully-para cooling rate approaches $\mathcal{O}(1)$ at such high temperatures \citep{Glover2008},
so the total cooling rate is insensitive to the actual ortho-para ratio.

However, the assumption that radiative transitions remain irrelevant is heavily constraining. For an optically-thin gas (so that we are only concerned with the photon-emitting transition), the radiative rate contribution can be ignored so long as $\Gamma_{O\rightarrow P,{\rm DM}} \le n_X \gamma_{X, O\rightarrow P, {\rm DM}}$, for dominant collision species $X$. Then, assuming protons are dominant, for a given $T$, we have
\begin{align}
    \Gamma_{O\rightarrow P,{\rm DM}} \le n_X \gamma_{X, O\rightarrow P, {\rm DM}}(T) \quad \to \quad
    \racx{9}{8}{-7} \Gamma_{O\rightarrow P, {\rm SM}} \le n_p \racx{-1}{-3/2}{-1/2} \gamma_{p, O\rightarrow P, {\rm SM}}(\tilde{T}_r)
\end{align}
or 
\begin{equation}
    \frac{\ra{10}\rc{19/2}}{\rx{-13/2}} \le \num{.5e21} \left(\frac{n_p \gamma_{p, O\rightarrow P,{\rm SM}}(\tilde{T}_r)}{\SI{1}{\per\second}}\right).
\end{equation}

This condition holds for the example cases given in the main article, but can be violated relatively easily, for example $(M, m, \alpha) = (\SI{1}{\giga\electronvolt},\SI{5.11}{\mega\electronvolt},0.02)$, $\tilde{T}_r\le\SI{e4}{\kelvin}$ and $n_p=n_{\rH}/\num{e4}=\SI{1}{\per\centi\meter\cubed}$. 

\section{Re-scaling combined cooling rate}\label{app:rescale_Ctot}
The literature on molecular hydrogen cooling rates tends to present the low-density limit as a single analytic fit of the full experimental or numerical rovibrational results, ${\cal C}_{\rm rovib}$ (see e.g. \citet{Galli1998,Glover2008}). This presents a difficulty in applying the re-scaling procedure presented here because the cooling process is composed of different physical processes, apparently complicating our assumption of a single energy scale by which the process can be non-dimensionalized and the fit re-scaled (a problem not unique to molecular cooling). Here, we show how these complications can be resolved by applying piecewise re-scalings in the temperature regime relevant to each physical process. 

The re-scaling of the temperature regimes themselves may lead to a gap between the rescaled curves. In this case, the higher temperature process is strongly suppressed below threshold and dominates above threshold, so 
we extrapolate each re-scaling out to the re-scaled threshold temperature and allow these curves to meet with a discontinuous derivative. 

For the case of $n\rightarrow0$ molecular cooling, there are only two processes to consider: rotational and vibrational level transitions. The net rate is the simple sum, ${\cal C}_{\rm tot}={\cal C}_{\rm rot}+{\cal C}_{\rm vib}$. Further, rotational transitions dominate at low temperatures (insufficient energy to induce a vibrational transition) and vibrational transitions dominate at high temperatures (larger energy transition), with a clear changeover between the regimes. Defining the pivot temperature $T_0$ as the temperature where the rotational cooling rate equals the vibrational cooling rate, we can divide the total rovibrational rate into
\begin{equation}
	{\cal C}_{\rm rovib}(T)={\cal C}_{{\rm rovib},T<T_0}(T)+{\cal C}_{{\rm rovib},T\ge T_0}(T) = R(T) + V(T).
\end{equation}
Note that $R(T)$ is only defined on the interval $0<T\le T_0$ and $V(T)$ is only defined on the interval $T_0\le T<\infty$.

From Section \ref{sec:cooling}, we know $R$ and $V$ can be re-scaled as , 
\begin{align}
    R_{\rm DM}(T) &= g(\ra{},\rc{},\rx{}) R_{\rm SM}(\tilde{T}_r) \\
    V_{\rm DM}(T) &= g(\ra{},\rc{},\rx{}) V_{\rm SM}(\tilde{T}_v).
\end{align}
But, since $R$ and $V$ are only defined for $T\le T_0$ and $T\ge T_0$, $R_{\rm DM}$ is only defined for $T\le \left((\rac{2}{2})/\rx{}\right) T_0=T_{0,r}$ and $V_{\rm DM}$ is only defined for $T\ge \left((\rac{2}{3/2})/\rx{1/2}\right) T_0=T_{0,v}$. Thus, the sum of $R_{\rm DM}$ and $V_{\rm DM}$ only gives the re-scaled rovibrational cooling rate across all $T$ if $T_{0,r}\ge T_{0,v}$ (effectively $\rx{}/\rc{}\le 1$), i.e.
\begin{equation}
	{\cal C}_{\rm rovib,DM}(T) = R_{\rm DM}(T) + V_{\rm DM}(T) \iff \rx{}\le \rc{}.
\end{equation}

The other case, $\rx{} > \rc{}$, requires additional information to fill the two gaps, $T_{0,r}<T\le T_{0,\rm DM}$ and $T_{0,\rm DM}\le T<T_{0,v}$, where $T_{0,\rm DM}$ is the re-scaled pivot point. One option exploits the high (low) temperature behavior of the rotational (vibrational) analytic curves from  \citet{Hollenbach1979}. Essentially, the rotational and vibrational curves both behave logarithmically at the temperature extremes, so a linear extrapolation of $R_{\rm DM}$ from $T_{0,r}$ to $T_{0,\rm DM}$ and of $V_{\rm DM}$ from $T_{0,v}$ to $T_{0,\rm DM}$ provides an acceptable fit.\footnote{We define linear extrapolation here as fitting the curve of interest, defined on some interval, to a function of the form $y= a x + b$ and evaluating the function outside the interval boundary.} The final full rovibrational cooling curve is given by
\begin{equation}
	{\cal C}_{\rm rovib,DM}(T) = 
	\begin{cases}
		R_{\rm DM}(T)+V_{\rm DM}(T) & \rx{} \le \rc{} \\
		R_{\rm DM}(T) & \rx{} > \rc{} \;\&\; T\le T_{0,r} \\
		{\rm linExtrap}\left(R_{\rm DM}(T)\right) & \rx{} > \rc{} \;\&\; T_{0,r}<T<T_{0,\rm DM} \\
		{\rm linExtrap}\left(V_{\rm DM}(T)\right) & \rx{} > \rc{} \;\&\; T_{0,\rm DM}\le T<T_{0,v} \\
		V_{\rm DM}(T) & \rx{} > \rc{} \;\&\; T_{0,v}\le T \\
	\end{cases},
\end{equation}
where ${\rm linExtrap}(F(T))$ is the linear extrapolation of the function $F(T)$. This is the approach taken in Figures \ref{fig:ldl_scaled_cooling} and \ref{fig:total_cooling_curve}. Other possibilities include different extrapolation functions or simply using the Hollenbach and McKee rates in those regions.

Using the rotational and vibrational rates from \citet{Hollenbach1979}, we find for $\rHt-\rH$ collisions that $T_0=\SI{856}{\kelvin}$ and $T_{0,\rm DM}\approx \left[\racx{2}{3/2}{-0.54}\right] \; T_{0,\rm SM}$ across the parameter space used in Figures \ref{fig:ldl_scaled_cooling} and \ref{fig:total_cooling_curve}. For $\rHt-\rHt$ collisions, we find $T_0=\SI{5.40e3}{\kelvin}$ and $T_{0,\rm DM}\approx \left[\racx{2}{1.58}{-0.586}\right]\; T_{0,\rm SM}$. We have been unable to determine an appropriate pivot temperature and corresponding re-scaling for the $\rHt-\{e,p\}$ collisions as of publication. As such, we will use rotational scaling for the entire temperature range, equivalent to $T_0\rightarrow\infty$. See Section \ref{sec:other_species} for more information.

\bibliographystyle{aasjournal}
\bibliography{molecular}

\end{document}